\documentclass[superscriptaddress, twocolumn, prx, nofootinbib]{revtex4-1}
\usepackage[utf8]{inputenc}
\usepackage{graphicx}
\usepackage{amsmath}
\usepackage{amsfonts}
\usepackage[table]{xcolor}
\usepackage[caption=false]{subfig}
\usepackage{array}
\usepackage{floatrow}
\usepackage{pbox}
\usepackage{physics}
\definecolor{linkcolor}{rgb}{0,0,0.6} 
\usepackage[pdftex,colorlinks=true,
	pdfstartview = FitV,
	linkcolor    = linkcolor,
	citecolor    = linkcolor,
	urlcolor     = linkcolor,	
	hyperindex   = true,
	hyperfigures = false]{hyperref}

\newcommand{\mrd}{\mathrm d}
\newcommand{\mre}{\mathrm e}
\newcommand{\mean}[1]{\left\langle #1 \right\rangle}

\newcommand{\fe}{f_\mathrm{ex}}
\newcommand{\fint}{f_\mathrm{int}}
\newcommand{\fa}{f_\mathrm{ac}}
\newcommand{\fs}{f_\mathrm{stall}}
\newcommand{\Po}{P_\mathrm{ex}}
\newcommand{\Pa}{P_\mathrm{ac}}
\newcommand{\Pc}{P_\mathrm{ch}}
\newcommand{\tot}{_\mathrm{tot}}
\newcommand{\therm}{\mathrm{th}}
\newcommand{\td}{\mathrm{td}}
\newcommand{\ch}{\mathrm{ch}}
\newcommand{\ac}{_\mathrm{ac}}

\renewcommand{\vec}{\boldsymbol}

\begin{document}
\title{Autonomous engines driven by active matter: Energetics and
  design principles}

\author{Patrick Pietzonka}
\affiliation{DAMTP, Centre for Mathematical Sciences, University of Cambridge, Wilberforce Road, Cambridge CB3 0WA, United Kingdom}
\author{\'Etienne Fodor}
\affiliation{DAMTP, Centre for Mathematical Sciences, University of
  Cambridge, Wilberforce Road, Cambridge CB3 0WA, United Kingdom}
\author{Christoph Lohrmann}
 \affiliation{ II. Institut f\"ur Theoretische Physik, Universit\"at Stuttgart,
   70550 Stuttgart, Germany}
\author{Michael E. Cates}
\affiliation{DAMTP, Centre for Mathematical Sciences, University of
  Cambridge, Wilberforce Road, Cambridge CB3 0WA, United Kingdom}
\author{Udo Seifert}
 \affiliation{ II. Institut f\"ur Theoretische Physik, Universit\"at Stuttgart,
   70550 Stuttgart, Germany}

\date{\today}

\parskip 1mm

\begin{abstract}

  Because of its nonequilibrium character, active matter in a steady
  state can drive engines that autonomously deliver work against a
  constant mechanical force or torque. As a generic model for such an
  engine, we consider systems that contain one or several active
  components and a single passive one that is asymmetric in its
  geometrical shape or its interactions. Generally, one expects that
  such an asymmetry leads to a persistent, directed current in the
  passive component, which can be used for the extraction of work. We
  validate this expectation for a minimal model consisting of an
  active and a passive particle on a one-dimensional lattice. It leads
  us to identify thermodynamically consistent measures for the
  efficiency of the conversion of isotropic activity to directed
  work. For systems with continuous degrees of freedom, work cannot be
  extracted using a one-dimensional geometry under quite general
  conditions. In contrast, we put forward two-dimensional shapes of a
  movable passive obstacle that are best suited for the extraction of
  work, which we compare with analytical results for an idealised
  work-extraction mechanism. For a setting with many noninteracting
  active particles, we use a mean-field approach to calculate the
  power and the efficiency, which we validate by
  simulations. Surprisingly, this approach reveals that the
  interaction with the passive obstacle can mediate cooperativity
  between otherwise noninteracting active particles, which enhances
  the extracted power per active particle significantly.

\end{abstract}


\maketitle

\section{Introduction}

The concept of thermal equilibrium allows for a comprehensive
characterisation of passive many-body systems in terms of
thermodynamic key quantities such as entropy and temperature.  Through
the second law of thermodynamics, changes in these quantities
constrain the amount of work an external operator can extract when
forcing the system to undergo a
transformation~\cite{Callen}. In contrast, active matter offers
a class of systems that goes beyond the scope of these well
established concepts. Such systems typically comprise an assembly of
self-driven components which operate far from thermal equilibrium by
extracting energy from their environment
\cite{rama10,cate12,marc13,bech16,zoet16}. Experimental realisations range
from swarms of bacteria~\cite{Kessler2004, Goldstein2007,
  Aranson2012} and assemblies of motile filaments~\cite{Dogic2012,
  Dogic2015} and of living cells~\cite{Ladoux2017, Sano2017} to
interacting Janus particles in a fuel bath~\cite{Golestanian2007,
  Palacci2013, Bechinger2013}. Phenomenological properties of active
matter, such as the emergence of clustering~\cite{Tailleur2008,
  Cates2015}, have been reproduced with simple mathematical models,
which can be either particle-based descriptions~\cite{Fily2012,
  Redner2013, Speck2013} or active field
theories~\cite{sten13, Wittkowski2014, Speck2014, nard17}.

Historically, key quantities of equilibrium thermodynamics
had been identified operationally through the interaction of the system
with embedded probes such as barometers and thermometers.
For a thermodynamic characterisation of active matter, several works
have followed this strategy. Extended
definitions of pressure~\cite{Marchetti2014, Brady2014, solo15b,
  Solon2018,spec16} and of chemical potential~\cite{Paliwal2018, Bertin2018}
have been proposed in active matter; moreover, a frequency-dependent
temperature has been introduced based on the violation of equilibrium
relations~\cite{Wilhelm2008, Visco2015, diet15a,Turlier2016,wulf17,
  Ahmed2018}. 

An important quest in the development of classical
thermodynamics was the formulation of fundamental design
principles for heat engines, which led to Carnot's statement of
the second law of thermodynamics~\cite{Carnot}.  For systems on small
scales, which are affected by ubiquitous thermal noise, stochastic
energetics~\cite{seki10} and stochastic thermodynamics~\cite{seif12}
provide a consistent framework for the generalisation of thermodynamic
concepts, such as the work that is either transferred to or extracted
from a nonequilibrium system. 

\begin{figure}
  \centering
  \includegraphics[width=\textwidth]{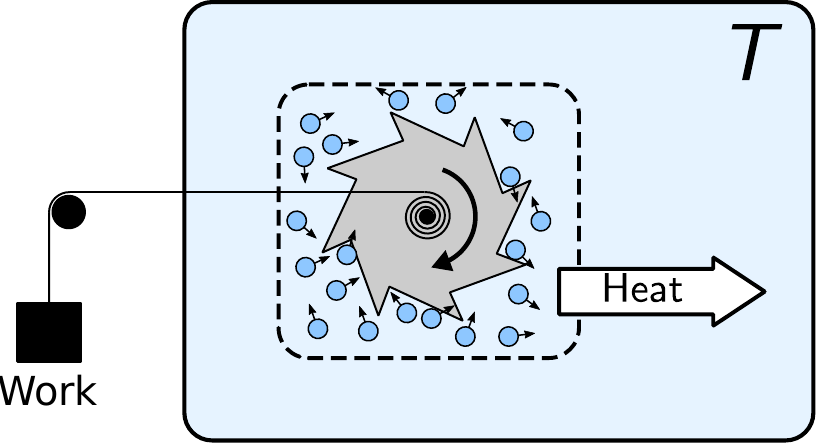}
  \caption{Schematic representation of an autonomous engine driven by
    active matter. An asymmetric cog-shaped passive obstacle (grey)
    rotates persistently in contact with a bath of active particles
    (dark blue). Applying an external load opposed to the spontaneous
    rotation, for instance by connecting the rotation axis to an
    external weight, the system produces work by lifting the
    weight. Besides, both the obstacle and the active particles are in
    contact with a thermostat at fixed temperature $T$, so that heat
    is constantly dissipated in the energy reservoir (light blue).  }
  \label{fig:extractor}
\end{figure}

Inspired by colloidal heat engines in a
thermal bath~\cite{schm08,blic12,Martinez2016}, the work delivered by
cyclic engines in contact with active baths has recently been
investigated both experimentally~\cite{kris16} and
theoretically~\cite{zaki17,mart18}. Such cyclic engines require an
external operator applying transformations according to some
time-periodic protocol.  An even simpler setting for nonequilibrium
systems that deliver work autonomously builds on ratchet
models~\cite{juel97,reim02}. They produce a persistent current in one
degree of freedom by rectifying fluctuations with some
asymmetric potential. In recent years, such ratchet models have been
used to illustrate nonequilibrium aspects of active
matter~\cite{gala07,wan08,tail09,ange10,kais12,ai13,kais13,kais14,koum14,ghos13,bech16,sten16,Reichhardt2017}.
Particularly inspiring are experiments where asymmetric cog-shaped
obstacles immersed in a bacterial bath autonomously undergo persistent
rotation~\cite{soko09,dile10,vizs17}---an observation that would be
prohibited in an equilibrium system due to time-reversal symmetry.  By
applying a sufficiently small countertorque to the rotor, its
rotation can be exploited for the extraction of mechanical work, as
illustrated in Fig.~\ref{fig:extractor}.

Despite the increasing development of such experiments, it remains to
evaluate and to rationalise the efficiency of such autonomous engines,
which should properly compare the extracted work with the dissipated
heat. In that respect, several studies strive to identify and to
quantify dissipation in simple models of active
particles~\cite{marc17,piet17b,gasp17,spec18,dabe18}, in relation with
entropy production and the irreversibility of the
dynamics~\cite{fodo16,mand17,pugl17,shan18}. These recent advances motivate a
systematic study of the performances of engines that extract work from
an active bath. What is the best shape of obstacles for delivering
optimal performance as an engine? How does one tune the properties of the
bath to extract maximum work? Answering such questions promises to
reveal new links between macroscopic observables and the
nonequilibrium character of active matter.

In this paper, we propose a consistent thermodynamic framework for
engines delivering work while being powered by active matter. En
route, we relate the extracted power to the energetics of the
self-propulsion of active particles. In defining the efficiency of the
work extraction, we distinguish between a fully detailed, microscopic
viewpoint and a more practical, coarse-grained viewpoint. The relevant
thermodynamic quantities can be identified in a simple lattice model
as well as in a general Langevin description of active Brownian
particles in continuous space.

We consider specific realisations of models for one or several active
particles interacting with a passive asymmetric obstacle that can move
in one linear direction against an external force. In each of these
models, we evaluate the power and efficiency of work extraction. In a
two-dimensional setting, the optimisation of these quantities leads to
nontrivial shapes of passive obstacles, which perform significantly
better than a simple chevron shape that has so far been a popular
model for ratchet effects in active
matter~\cite{gala07,wan08,tail09,ange10,kais12,kais13,kais14,sten16}.
For the case of many noninteracting active particles, we devise a
dynamical mean-field approach. It reveals, somewhat surprisingly, that
the extracted power per active particle is larger for many active
particles than for a single active particle. Numerical simulations of
the full many-body dynamics confirm these theoretical predictions.

The paper is organised as follows. In Sec.~\ref{sec:lattice}, we begin
with a simple lattice model and set up definitions concerning the
energetics, which are illustrated on the basis of exact
results. Section~\ref{sec:contdof} introduces the energetics for a
general Langevin description of active particles interacting with each
other and with a passive obstacle. Moreover, we derive a no-go
theorem, which excludes the possibility to extract work for an overly
simple class of obstacles. In Sec.~\ref{sec:single}, we calculate the
power and efficiency for a single active particle and discuss design
principles for the shape of a passive obstacle. Section~\ref{sec:many}
generalises to many noninteracting active particles, introducing our
dynamic mean-field theory, which is validated using numerical
simulations. We conclude in Sec.~\ref{sec:conclusions}.

\section{Minimal lattice model} 
\label{sec:lattice}
\subsection{Setup} 

\begin{figure}
  \centering
  \includegraphics[scale=0.75]{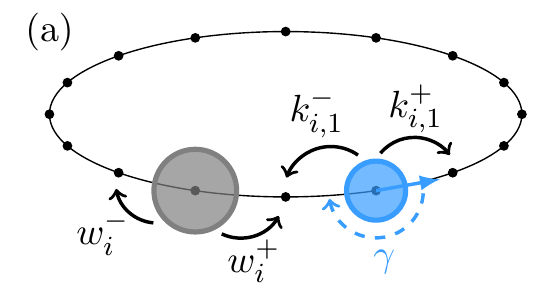}
  \includegraphics[scale=0.75]{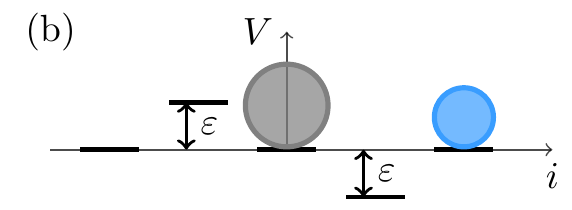}
  \includegraphics[scale=0.5]{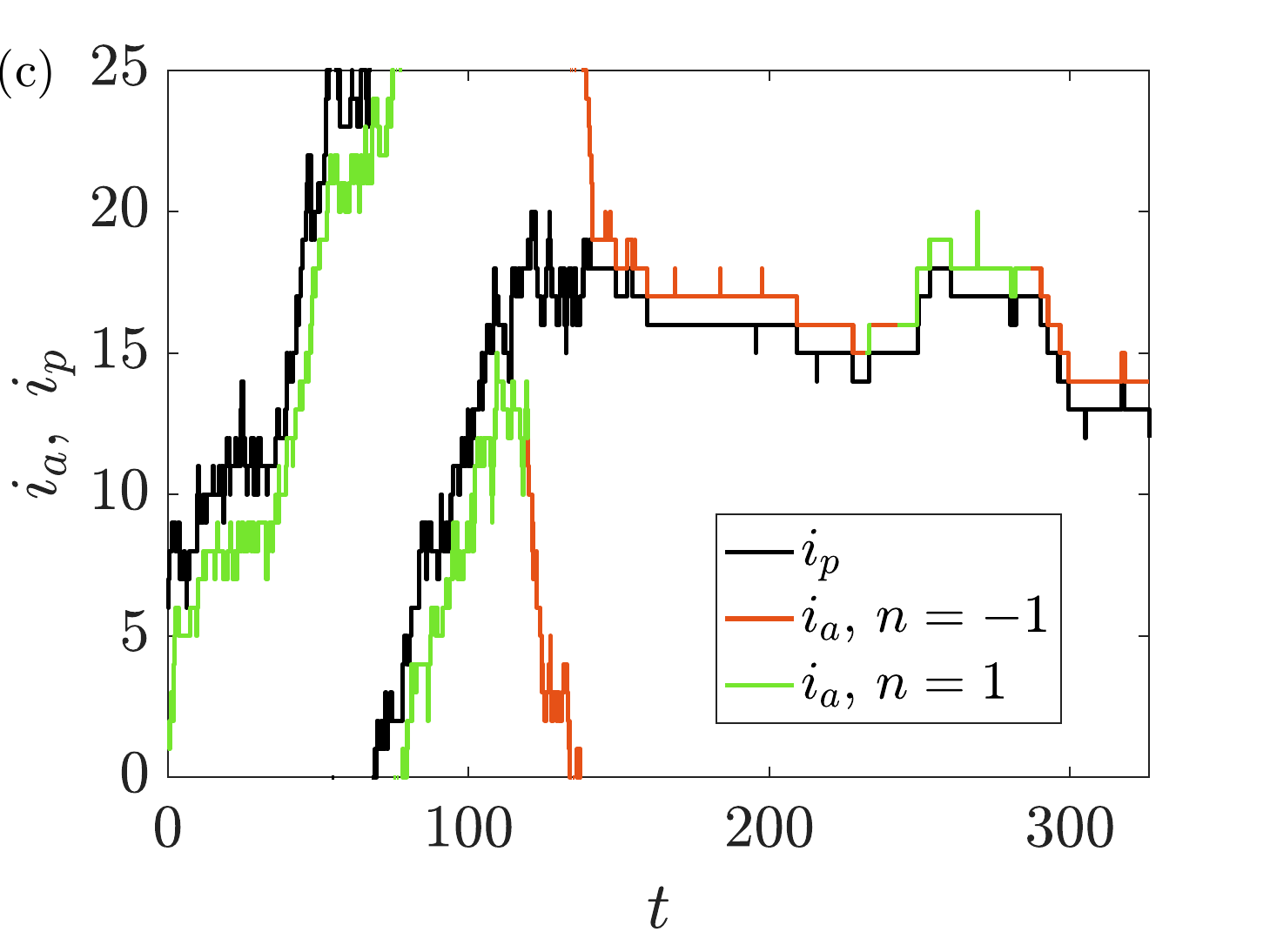}

  \caption{(a) Setup for the lattice model of an active particle
    (blue) interacting with a passive one (gray) in a periodic
    geometry. (b) The particles interact via on-site exclusion and via
    an asymmetric potential ranging to the next lattice site. (c)
    Sample trajectories for the passive particle (black line) and the
    active particle (green or red line, depending on the internal degree of
    freedom). The active particle first pushes the passive one from
    behind, until its internal degree of freedom changes
    direction at $t\simeq 120$. Then, after circuiting once around the ring, it hits the passive particle on the attractive
    other side and sticks to it for the remainder of the shown time
    interval.  Parameters: $L=25$, $k_0=1$, $\fa=1$, $w_0=1$,
    $\gamma=0.03$, $\varepsilon=5$, and $\fe=0.1$.}
  \label{fig:latticemodel}
\end{figure}

Lattice models have repeatedly been used as minimal models for the
analysis of various aspects of active matter~\cite{thom11,solo13,slow16,piet17b,whit18}.
For our purpose of studying the extraction of work, we consider a one-dimensional lattice with $L$
sites, periodic boundary conditions, and one active and one passive
particle, as shown in Fig.~\ref{fig:latticemodel}(a). Both particles can
hop to unoccupied neighbouring lattice sites. The positions of the
particles at time $t$ are denoted as $i_a(t)$ and $i_p(t)$ for the
active and passive particle, respectively.  We define the signed distance
between the two particles as
\begin{equation}
  \label{eq:reldisc}
  i(t)\equiv [i_a(t)-i_p(t)]\mod L
\end{equation}
where the modulo operation is due to the periodic boundaries and is
applied such that $1\leq i(t)<L$.  The free active particle is
modelled as a run-and-tumble particle that has an internal degree of
freedom~$n(t)$ that switches stochastically between $+1$ and $-1$ at a
Markovian rate~$\gamma$. The inverse of this rate $1/\gamma$
quantifies the persistence time of the active particle. According to
the state of the variable~$n$, the particle hops preferentially to the
right or to the left, as detailed below. The free passive particle has
hopping rates that are biased in the direction of the applied external
force. In order to obtain a persistent, directed current even in the
absence of the external force, the left-right symmetry of the system
needs to be broken. In the setting we ultimately have in mind, this
symmetry breaking would be achieved by giving the passive particle
some asymmetric shape, which, however, is not possible in one dimension.
As a simple way to still break the symmetry while preserving the
dynamics of the free particles, we introduce, in addition to the
hard-core exclusion of the particles, an asymmetric short-range
interaction potential
\begin{equation}
  V_i\equiv -\varepsilon(\delta_{i,1}-\delta_{i,L-1})
\end{equation}
for $1\leq i<L$, as shown in Fig.~\ref{fig:latticemodel}(b). Thus, the
passive particle attracts the active one, if the latter is one lattice
site to the right. Conversely, the passive particle repels the active
one, if the latter is one lattice site to the left. This local
interaction mimics the effect that some notch in the shape of the
passive obstacle would have, which traps the active particle on one
side but not on the other one.

The dynamics of the system is modelled as a continuous-time Markov
process. For the identification of physical heat, we require this
process to be thermodynamically consistent, which constrains the
transition rates depending on the driving forces and the potential~\cite{seif12}. The passive particle
hops to the right or left at rates $w_i^+$ and $w_i^-$, respectively,
which depend on the distance $i(t)$. In contact with a heat bath
at a constant temperature, these rates are constrained
by the local detailed balance condition
$w_i^+/w_{i-1}^-=\exp(-\fe+V_i-V_{i-1})$, with the external force
$\fe$ acting on the passive particle in negative direction. In this section, we set the lattice
spacing, the temperature, and Boltzmann's constant to one. The local
detailed balance condition is satisfied by setting
\begin{equation}
  w_i^\pm=w_0\mre^{(\mp\fe+V_i-V_{i\mp1})/2}
  \label{eq:rate_w}
\end{equation}
for all transitions that do not lead
to an overlap of the particles. The prefactor $w_0$ is a rate of
reference that is not constrained by thermodynamics, determining the
diffusivity of the passive particle. 
The hard-core exclusion is accounted
for by setting all rates involving the state $i=0$ (or, equivalently,
$i=L+1$) to zero.

The active particle is endowed with a self-propulsion mechanism that
allows for chemically driven translational transitions biased towards the direction
given by the internal degree of freedom $n(t)$. Nonetheless, the transitions
can also be induced by the passive influence of thermal noise and
potential forces. A minimal thermodynamically consistent model
accounts for both types of transition~\cite{piet17b}. It ascribes to
the thermal transitions of the active particle the transition rates
\begin{equation}
  k_{i,\therm}^\pm=k_{0,\therm}\mre^{(V_i-V_{i\pm1})/2}
  \label{eq:rate_kp}
\end{equation}
with a rate of reference $k_{0,\therm}$ The chemically driven
transitions occur at rates
\begin{equation}
  k_{i,n,\ch}^\pm=k_{0,\ch}\mre^{(\pm n \varDelta\mu+V_i-V_{i\pm1})/2}
  \label{eq:rate_ka}
\end{equation}
with another rate of reference $k_{0,\ch}$ and the chemical free energy
$\varDelta\mu$ that is transduced in a transition parallel to the
preferred (active) jump direction $n$. On
a mesoscopic scale, where information on the microscopic chemical
process is not accessible, the two types of transition cannot be
distinguished, leading to the combined rates
\begin{subequations}
\label{eq:rate_kdef}
\begin{align}
  k_{i,n}^\pm&\equiv k_{i,\therm}^\pm+k_{i,n,\ch}^\pm\\
&= k_0\mre^{(\pm n \fa+V_i-V_{i\pm1})/2}.
  \label{eq:rate_k}
\end{align}
\end{subequations}
In Eq.~\eqref{eq:rate_k}, we have brought these combined rates to the form of a local detailed balance
relation, with 
an effective, ``active'' force
\begin{equation}
  \fa\equiv\ln\frac{k_{0,\therm}+k_{0,\ch}\mre^{\varDelta\mu/2}}{k_{0,\therm}+k_{0,\ch}\mre^{-\varDelta\mu/2}}
\end{equation}
and
\begin{equation}
  k_0\equiv\sqrt{k_{0,\therm}^2+k_{0,\ch}^2+2k_{0,\therm}k_{0,\ch}\cosh(\varDelta\mu/2)}.
\end{equation}
Even though the active force does not enter the microscopic
rates~\eqref{eq:rate_kp}, it emerges as a useful quantity for a
dynamical, mesoscopic description of the active particle as being pulled
by a fictitious external force $\fa$ acting in the direction of
$n(t)$. Unlike the microscopic parameter $\varDelta\mu$, the active
force can be inferred on a mesoscopic scale, for example through the
force that is required to stall an active particle with persistently
positive $n$. This property allows us to define the active force 
independently of the microscopic dynamics, which would typically be
much more complex than what is captured by the minimal model used here.

In the state space spanned by the variables $(i_p,i_a,n)$, the
transition rates \eqref{eq:rate_w}, \eqref{eq:rate_kdef}, and $\gamma$
give rise to a stochastic dynamics, that is illustrated in
Fig.~\ref{fig:latticemodel}(c) with a sample trajectory. The
corresponding master equation leads to a stationary probability
distribution $p(i_p, i_a,n)$. Because of the translational symmetry of the
total system, this distribution can be written as
\begin{equation}
  p(i_p,i_a,n)=p(i,n)/L,
\end{equation}
where $p(i,n)$ is the stationary distribution on the state space
spanned by $n$ and the relative coordinate
\eqref{eq:reldisc}. Combining the transition rates that increase and
decrease $i$, we find for this distribution the reduced
stationary master equation
\begin{align}
  0&=\mrd p(i,n)/\mrd t\nonumber\\&=[w_{i+1}^++k_{i+1,n}^-]p(i+1,n)+[w_{i-1}^-+k_{i-1,n}^+]p(i-1,n)\nonumber\\
   &\qquad+\gamma p(i,-n)-[w_i^++w_i^-+k_{i,n}^++k_{i,n}^-+\gamma]
     p(i,n).
     \label{eq:master}
\end{align}
For finite $L$, this system of equations with $2(L-1)$ unknowns can be
solved using linear algebra.

\subsection{Energetics}
With the steady-state distribution $p(i,n)$ at hand, the
total particle current $J$, or average velocity of the particles, is given
by
\begin{equation}
  \label{eq:disc_current}
  J=\sum_{i,n}p(i,n)[w_i^+-w_i^-]=\sum_{i,n}p(i,n)[k_{i,n}^+-k_{i,n}^-].
\end{equation}
Since the particles cannot pass through each other, the average
velocities of the passive and the active particle must be the same, which
leads to the second equality.

The extracted power is the rate of work performed against the
force $\fe$:
\begin{equation}
  \label{eq:pout_def}
  \Po\equiv J\fe.
\end{equation}
This power is positive when the external force acts in the direction
opposite to the current, while $\Po<0$ means that work
is performed on the system. To extract positive work, the
external force must be nonzero and opposite to the direction of the
current at zero force, and its absolute value must be smaller than the
stall force $f_\mathrm{stall}$ at which the current vanishes.

The input of chemical work into the total system
stems from the chemically powered, active transitions of the active particle
\begin{equation}
  \label{eq:pch_def}
  \Pc\equiv\sum_{i,n}\varDelta\mu\, n\,p(i,n) [k_{i,n,\ch}^+-k_{i,n,\ch}^-].
\end{equation}
On the other hand, the total rate of entropy production follows from its standard
definition~\cite{schn76} as
\begin{align}
  \sigma\tot&\equiv\sum_{i,n}[p(i,n)k_{i,n,\ch}^+-p(i+1,n)k_{i+1,n,\ch}^-]\ln\frac{k_{i,n,\ch}^+}{k_{i+1,n,\ch}^-}\nonumber\\
              &\qquad+\sum_{i,n}[p(i,n)k_{i,\therm}^+-p(i+1,n)k_{i+1,\therm}^-]\ln\frac{k_{i,\therm}^+}{k_{i+1,\therm}^-}\nonumber\\
            &\qquad+\sum_{i,n}[p(i,n)w_{i}^+-p(i-1,n)w_{i-1}^-]\ln\frac{w_{i}^+}{w_{i-1}^-}\nonumber\\
  &=\Pc-\Po\geq 0.
  \label{eq:stot_disc}
\end{align}
Thus, the chemical power $\Pc=\Po+\sigma\tot$ is transferred to both
extracted power~$\Po$ and dissipated power~$\sigma\tot$~\cite{parm99}. The latter is
essentially the heat that is dissipated into the environment but may
also include the change of entropy in chemical
reservoirs~\cite{seif11}.  The thermodynamic efficiency associated
with the extraction of work can be defined as
\begin{equation}
  \label{eq:eta_th}
  \eta_\td\equiv\frac{\Po}{\Pc}.
\end{equation}

On the mesoscopic scale, where active and passive transitions of the
active particle are typically indistinguishable, an exact evaluation
of $\sigma\tot$ based on observations is not possible. Hence, a
coarse-grained thermodynamic approach is necessary to evaluate the
performance of the engine. Applying the concepts of stochastic
thermodynamics to the coarse-grained model involving the combined
transition rates $k_{i,n}^\pm$ yields the coarse-grained entropy
production rate
\begin{align}
  \varSigma&\equiv\sum_{i,n}[p(i,n)k_{i,n}^+-p(i+1,n)k_{i+1,n}^-]\ln\frac{k_{i,n}^+}{k_{i+1,n}^-}\nonumber\\
            &\qquad+\sum_{i,n}[p(i,n)w_{i}^+-p(i-1,n)w_{i-1}^-]\ln\frac{w_{i}^+}{w_{i-1}^-}\nonumber\\
&=\Pa-\Po.
  \label{eq:cgsigma}
\end{align}
This rate of entropy production is also obtained by comparing the
forward and time-reversed path probabilities for $i_p$ and $i_a$
without taking into account the types of transition for $i_a$.  We
identify the ``active power''
\begin{equation}
  \label{eq:pac_def}
  \Pa\equiv\sum_{i,n}\fa n\,p(i,n) [k_{i,n}^+-k_{i,n}^-]
\end{equation}
as the rate of work performed by the fictitious active force~$\fa$.
Since the coarse-grained entropy production satisfies
$0\leq\varSigma\leq\sigma\tot$, we find
\begin{equation}
  \Pc\geq\Pa\geq\Po.
  \label{eq:Porder}
\end{equation}
Hence, the active power gives a stronger bound on the extracted
power than the full chemical power. Since this inequality shows that the difference between
$\Pc$ and $\Pa$ is inevitably dissipated into the environment, we ask in the following how
much of $\Pa$ can be extracted as useful work $\Po$ and define the
``active efficiency''
\begin{equation}
  \label{eq:eta_ac}
  \eta\equiv\frac{\Po}{\Pa}=\frac{\Po}{\varSigma+\Po}=\frac{\fe}{\fa}\frac{\sum_{i,n}\,p(i,n) [k_{i,n}^+-k_{i,n}^-]}{\sum_{i,n} n\,p(i,n) [k_{i,n}^+-k_{i,n}^-]}.
\end{equation}
The identification of $\Po$ and $\Pa$ and their relation to $\Pc$
constitute the first main result of this paper. Importantly, the
former two are accessible on a mesoscopic scale, and, thus, retain their
significance beyond our specific minimal model for the chemical
driving process, as shown below for a description with
continuous degrees of freedom.  Consequently, we use $\Po$ and
$\eta$ as the main quantities of interest to characterise the
performance of an engine.

\begin{figure}
  \centering
  \includegraphics[width=\textwidth]{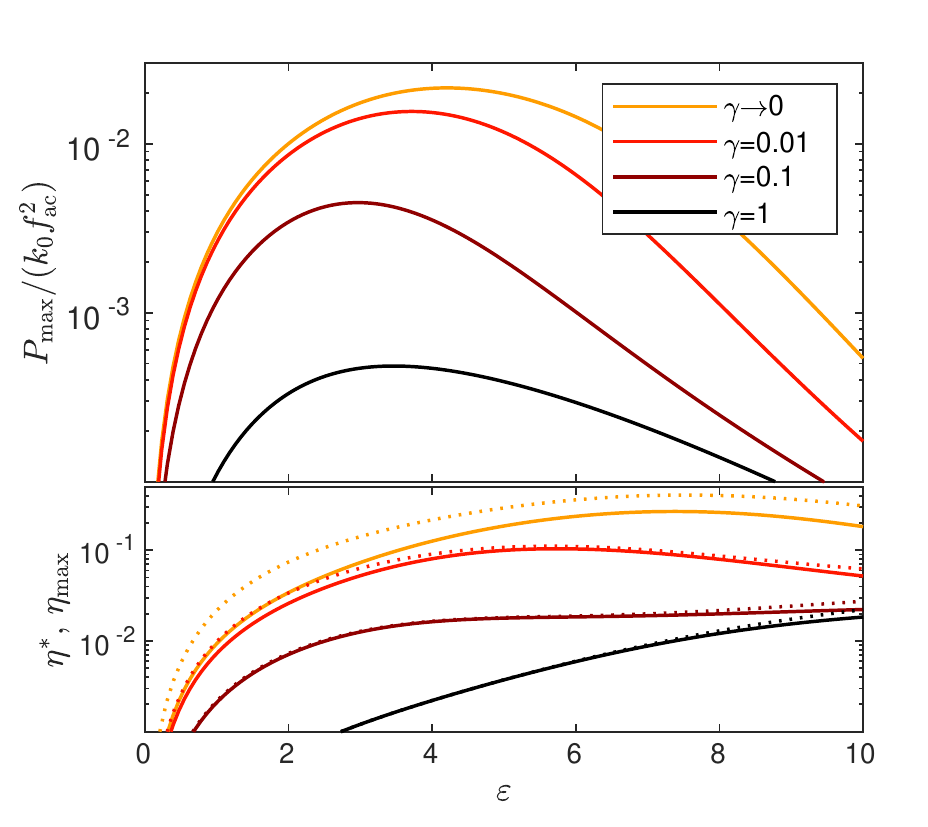}
  \caption{Numerical evaluation of the engine performance in the
    lattice model with $L=10$ sites. Top: The maximum extractable
    power as a function of the dimensionless interaction strength
    $\varepsilon$. The parameters of the active particle, $k_0=1$,
    $\fa=1$ and selected values of $\gamma$ are kept fixed, while the
    extracted power $\Po$ is optimised with respect to $w_0$ and
    $\fe$. The limit of small tumbling rate $\gamma\to 0$ (orange) is
    evaluated using the effective potential~\eqref{eq:Veff}. Bottom:
    The active efficiency $\eta^*$ at maximum power (solid lines) and
    the maximal active efficiency $\eta_\mathrm{max}$ (dotted lines)
    optimised again with respect to $w_0$ and $\fe$. }
  \label{fig:maxpow}
\end{figure}

We now explore the dependence of the power and
efficiency of work extraction on the various parameters of the lattice
model. For this purpose, the stationary master
equation~\eqref{eq:master} is solved numerically. As the main
parameter of interest, we choose the interaction strength~$\varepsilon$,
which represents a measure for the asymmetry of the passive work-extraction mechanism, and which is dimensionless in our units with
$k_BT=1$. For each combination of parameters shown in
Fig.~\ref{fig:maxpow}, we optimise the extracted power with respect to
the external force $\fe$ and the passive diffusivity $w_0$, leading to optimal
parameters $\fe^*$ and $w_0^*$. Evaluating $\eta$ at these parameters
gives the active efficiency at maximum output power $\eta^*$. In
addition, we calculate the global maximum of the efficiency
$\eta_\mathrm{max}$ over all values of $\fe$ and $w_0$, which turns
out to be only slightly larger than $\eta^*$.

In Appendix~\ref{sec:app_twostate} we present analytical calculations
for various limiting cases of the model. In particular, we find that
in the high persistence regime for $\gamma\ll w_0,k_0$, the power and
efficiency of work extraction generally become highest. The
corresponding curves in Fig.~\ref{fig:maxpow} saturate in the limit
$\gamma\to 0$. In contrast to that, the regime where the orientation
of the active particle switches very quickly resembles an equilibrium
system, where work cannot be extracted. Finally, another analytically
tractable limiting case is the one where the interaction is strong,
$\varepsilon\to\infty$. By adjusting the active and external forces,
the active efficiency can get arbitrarily close to one, showing that
there is no universal upper bound on the efficiency smaller than the
trivial bound $\eta\leq 1$.

\section{General theory  for continuous degrees of freedom} 
\label{sec:contdof}

\subsection{Setting and energetics}

In a general context, we consider a set of $N$ active particles
in a two- or three-dimensional channel or box with periodic boundary
conditions. These particles interact with each other and with a
passive object serving as the work extractor. The interactions are
mediated by pair potentials, and, for simplicity, we neglect
hydrodynamic interactions. The passive object, in the following
referred to as the ``obstacle'', has a fixed shape and is constrained to
move along a single degree of freedom against an external force. For
notational simplicity, we take this degree of freedom to be the
translational one associated with the direction $\vec{e}_x$ and keep
the orientation of the particle fixed. Thus, while the active
particles are \textit{a priori} free to move and rotate in any
direction, we keep the obstacle effectively fixed to a
one-dimensional ``railway line.''  The formulation of the model for a
rotating obstacle at a fixed position and subject to an
external torque, as illustrated in Fig.~\ref{fig:extractor}, would be
analogous.

We denote the positions of the active particles as $\vec{r}_a^i$, with $i$ labelling the
particle index, and the position of the passive obstacle as
$\vec{r}_p$. The dynamics of the latter is modelled
through the overdamped Langevin equation
\begin{equation}
  \dot{\vec{r}}_p=(\mu_p[-\fe+\sum_i\vec\nabla
  V(\vec{r}_a^i-\vec{r}_p)\cdot\vec{e}_x]+\zeta_p)\vec{e}_x.
  \label{eq:langevin_cont_p}
\end{equation}
Here, $\mu_p$ is the mobility of the passive obstacle (in the
$\vec{e}_x$ direction), $\fe$ is the external force applied in the
negative $\vec{e}_x$ direction and $V(\vec{r})$ is the interaction
pair potential between the obstacle and each of the active
particles, with $\vec\nabla$ acting on the distance vector
$\vec{r}=\vec{r}_a^i-\vec{r}_p$. The term $\zeta_p$ is Gaussian
white noise with correlations
$\mean{\zeta_p(t)\zeta_p(t')}=2D_p\delta(t-t')$ and the diffusion
coefficient $D_p=\mu_pk_BT$ at temperature $T$.

The active particles are chemically driven in the direction of their
internal orientation vectors $\vec{n}^i$. Their positions evolve
according to the overdamped Langevin equation
\begin{equation}
  \dot{\vec{r}}_a^i= u\ac\vec{n}^i+\vec{\mu}_a^i\vec
  f^i_\mathrm{pot}+\vec{\zeta}^i_a
  \label{eq:langevin_cont_a}
\end{equation}
with an active velocity $u\ac$ and the potential force
\begin{equation}
  \vec f_\mathrm{pot}^i\equiv-\vec\nabla V(\vec{r}_a^i-\vec{r}_p)- \sum_{j\neq i}\vec\nabla U(\vec{r}_a^i-\vec{r}_a^j).
\end{equation}
Here, $U(\vec{r})$ is the pair potential for interactions between
active particles, , with $\vec\nabla$ acting on the distance vector
$\vec{r}=\vec{r}_a^i-\vec{r}_a^j$.  We consider two microscopic
origins of the noise term, which we decompose as
$\vec{\zeta}_a^i=\vec{\zeta}^i_{\therm}+\zeta^i_\ch\vec{n}^i$. As
described in Ref.~\cite{piet17b}, these two terms arise in the
continuum limit of a lattice model analogous to the one in
Sec.~\ref{sec:lattice}. First, the thermally induced translational
Brownian motion of the active particle is modelled with an isotropic
noise term $\vec{\zeta}^i_{\therm}$ with correlations
$\mean{\vec{\zeta}^i_{\therm}(t)\otimes\vec{\zeta}^j_{\therm}(t')}=2D_{\therm}\delta(t-t')\delta_{ij}\vec{1}$.
Second, the noise in the chemical reaction couples to the driven
motion of the particle in the direction $\vec{n}^i$, which is
reflected in the one-component noise term $\zeta^i_\ch$ with
correlations
$\mean{\zeta^i_\ch(t)\zeta^j_\ch(t')}=2D_{\ch}\delta(t-t')\delta_{ij}$. The
fluctuation-dissipation theorem requires the mobility tensor to have
two components
$\vec{\mu}_a^i=\mu_\therm\vec{1}+\mu_\ch\vec{n}^i\otimes\vec{n}^i$,
according to the diffusion coefficients $D_\therm=\mu_\therm k_BT$ and
$D_\ch=\mu_\ch k_BT$.  The thermal mobility $\mu_\therm$ is given by
the inverse of the Stokes friction of the particle in the surrounding
fluid. The chemical mobility $\mu_\ch$ depends on the details of the
self-propulsion mechanism. For example, the continuum limit of the
discrete microscopic model discussed in Sec.~\ref{sec:lattice} yields
$\mu_\ch=u\ac d/\varDelta\mu$ for a small driving affinity
$\varDelta \mu\ll k_BT$ of a reaction event that comes with the
displacement $d$.

The vectors $\vec{n}^i$ perform isotropic rotational Brownian motion
on the unit circle or unit sphere with rotational diffusion
coefficient $D_r$. We take this rotational diffusion to be independent
of the position of the particles; i.e., we assume that there are no
alignment interactions among active particles or between active
particles and the obstacle. This aspect of the model has been
validated experimentally for at least one class of autophoretic Janus
colloids, whose orientation is indeed left unaffected upon contact
with an obstacle~\cite{volp11}. However, for rod-shaped active
particles or pusher- or puller-type microswimmers, steric or hydrodynamic
alignment interactions are present \cite{dile10,dres11} and would need
to be included in a more complex dynamics for $\vec{n}^i$. Moreover,
we note that active Ornstein-Uhlenbeck particles~\cite{fodo16} can
formally be implemented in the present formalism by allowing the
length of $\vec{n}^i$ to fluctuate as well, such that $\vec{n}^i$
performs an Ornstein-Uhlenbeck process. Finally, a setting with fixed
obstacle positions, commonly considered as active
ratchets~\cite{gala07,wan08,tail09,ange10,kais12,ai13,kais13,kais14,koum14,ghos13,bech16,sten16,Reichhardt2017},
can also be used to extract mechanical work by applying the external
force directly to the active particles rather than to a passive
tracer. Our discussion of the thermodynamics and the design principles
below can straightforwardly be extended to this case through a change
of the reference frame.

The dynamics described by the Langevin equations leads to a stationary
distribution $p(\{\vec{r}^i_a\},\{\vec{n}^i\},\vec{r}_p)$. The mean
velocity of the obstacle can be expressed as an average with
respect to this distribution as
\begin{equation}
  J\equiv\mean{\dot{\vec{r}}_p\cdot\vec{e}_x}=-\mu_p\fe+\mu_p\sum_i\mean{\vec\nabla
    V(\vec{r}_a^i-\vec{r}_p)\cdot\vec{e}_x},
  \label{eq:Jcont}
\end{equation}
which leads to the extracted power $\Po=\fe J$.
The rate of total thermodynamic entropy production in the steady state follows
through the same steps as in Ref.~\cite{piet17b} as 
\begin{equation}
  \sigma\tot=(\Pc-\Po)/T\geq 0,
\label{eq:stot_cont}
\end{equation}
with the chemical power
\begin{equation}
  \Pc=Nu\ac^2/\mu_\ch+u\ac\sum_i\mean{\vec n^i\cdot\vec{f}^i_\mathrm{pot}}.
  \label{eq:Pch_cont}
\end{equation}
Equation~\eqref{eq:stot_cont} is analogous to Eq.~\eqref{eq:stot_disc} for
the discrete model, where $T$ had been set to $1$.

On a mesoscopic scale, only the dynamics of $\vec{r}_p$,
$\vec{r}^i_a$  and $\vec{n}^i$ can be observed, while the two sources
of the noise become indistinguishable. While this fact prohibits an
exact evaluation of the chemical power, we can, as before for the
lattice model, identify an active power as a coarse-grained
quantification of the input of energy. It can be defined
model independently as the rate of work
\begin{equation}
  \Pa\equiv \fa\sum_i\mean{\vec{n}^i\cdot\dot{\vec{r}}_a^i}
  \label{eq:Pac_cont0}
\end{equation}
that is performed by an effective active force apparently providing
propulsion in the direction of
$\vec{n}^i$~\cite{fodo14,fodo16,cagn17,nemo19}. Such an active force
is commonly used \textit{ad hoc} in theoretical models for active
particles that discard the details of the self-propulsion
mechanism~\cite{hage11,Fily2012,Redner2013,magg15,fara15,fodo18a}.  It
can be determined phenomenologically as the force required to stall an
active particle with persistent director $\vec{n}$.  Experimentally,
the active force can also be measured for a single free active
particle that undergoes rotational and translational diffusion. For
this purpose one applies an external force and evaluates the
velocity of the active particle at times when the director $\vec{n}$
is antiparallel to the external force. When this velocity reaches zero
on average, the absolute value of the external force matches that of
the active force. For the specific model at hand, we have
$\fa\equiv u\ac/(\mu_\therm+\mu_\ch)$, and the active power can be written
using Eq.~\eqref{eq:langevin_cont_a} as
\begin{equation}
  \Pa=Nu\ac^2/(\mu_\therm+\mu_\ch)+u\ac\sum_i\mean{\vec
    n^i\cdot\vec{f}^i_\mathrm{pot}}.
  \label{eq:Pac_cont}
\end{equation}
The difference between the effective input $\Pa$ and the actual output
power $\Po$ leads to the definition of the coarse-grained entropy production
$\varSigma\equiv (\Pa-\Po)/T$. Several recent works quantify
irreversibility in active matter in terms of the ratio of forward
and backward path probabilities~\cite{fodo16,marc17,mand17,pugl17,shan18,dabe18}. 
The same entropy production $\varSigma$ emerges in such a framework by
applying the standard principles of stochastic thermodynamics to the joint trajectory of $\{\vec{r}^i_a\}$, $\{\vec{n}^i\}$
and $\vec{r}_p$ and considering both $\{\vec{n}^i\}$ and $\fe$ as even
under time reversal. Other choices for the set of variables and
time reversal are conceivable and have indeed been explored as
characterisations for the irreversibility, yet only this choice
yields the connection to the active power $\Pa$ that is useful in the
context of work extraction. The ensuing entropy production $\varSigma$
is positive and, as a result of the coarse-graining procedure, smaller
than the full entropy production $\sigma\tot$, yielding again the
order~\eqref{eq:Porder} for the chemical, active, and extracted
power.

In the following, we assume that the chemical contribution to the
mobility $\mu_\ch$ is much smaller than the thermal contribution
$\mu_\therm$. This assumption is justified from a microscopic
perspective, if the displacement $d$ of the active particle associated
with an individual reaction event is sufficiently small, such that,
for external or potential forces $f$ that lead to velocities
$\mu_\therm f$ of the order of $u\ac$, we have
$\mu_\ch/\mu_\therm\sim f d/\varDelta\mu\ll 1$ (see also
Ref.~\cite{spec18}). In fact, $\mu_\ch$ might even be on the order of
magnitude of a slight geometric anisotropy of the thermal mobility
tensor itself, prohibiting the inference of $\mu_\ch$ on a mesoscopic
scale.  Under the assumption of small $\mu_\ch$, the chemical
power~\eqref{eq:Pch_cont} is much larger than the active
power~\eqref{eq:Pac_cont0}. Since the latter bounds the extracted power
$\Po$ from above, the thermodynamic efficiency $\eta_\td=\Po/\Pc$ is small. With the dominating first term in Eq.~\eqref{eq:Pch_cont}
being constant, maximising $\Po$ for fixed $u\ac$ and $\mu_\ch$ leads
also to maximal $\eta_\td$.

Alternatively, we consider analogously to Eq.~\eqref{eq:eta_ac} the
active efficiency $\eta=\Po/\Pa$ as a mesoscopically accessible
characterisation of the performance of a work extractor after
subtracting the inevitable chemical losses. For the remainder of the
paper, we focus on the model of active particles with an isotropic
mobility tensor $\vec{\mu}_a=\mu_a\vec{1}$ and diffusion coefficient
$D_a=\mu_a k_BT$, which reproduces the dynamics of the model
considered above up to corrections of the order of $\mu_\ch/\mu_\therm$. The
active force is then simply given by $\fa=u\ac/\mu_a$, where both the
speed $u\ac$ and the mobility $\mu_a$ are straightforward to determine
experimentally. This ``active Brownian particle'' model, which
discards the chemical noise in the self-propelled motion but keeps
thermal diffusive noise in both the translational and angular sectors,
is standard in the literature, as reviewed, e.g., in Ref.~\cite{bech16}.

\subsection{A no-go theorem}
We first consider a single passive particle, serving as an obstacle,
and a single active particle along a one-dimensional continuous
coordinate. The active particle has a director $n$ that jumps between
$\pm1$ at a position-independent rate $\gamma$.  The Langevin
equations~\eqref{eq:langevin_cont_p} and~\eqref{eq:langevin_cont_a}
then reduce to
\begin{subequations}
\begin{align}
  \label{eq:langevin_p}\dot x_p&=\mu_p[-\fe+V'(x_a-x_p)]+\zeta_p\\
  \label{eq:langevin_a}\dot x_a&=\mu_a[n\fa-V'(x_a-x_p)]+\zeta_a
\end{align}
\label{eq:langevin_ap}
\end{subequations}
for the respective positions $x_a$ and $x_p$ of the active and passive
particles on a ring with $x_a,x_p\in[0,1)$ with periodic boundary
conditions and with independent one-dimensional noise terms
$\zeta_{a,p}$.  The interaction potential $V(x)$ is a function of the
relative coordinate $x= x_a-x_p$ (with the prime denoting the derivative
with respect to $x$), which consists of a hard-core exclusion and an
additional, asymmetric interaction. Despite the similarity to the
lattice model considered in Sec.~\ref{sec:lattice}, it is not possible
to produce a persistent current against the external force for any
potential $V(x)$, as we now show.

Because of the hard-core exclusion, the mean velocities of the active and
the passive particle are both equal to the overall current
$J=\mean{\dot x_a}=\mean{\dot x_p}$. By rescaling the Langevin
equations \eqref{eq:langevin_a} and \eqref{eq:langevin_p} with the
respective mobilities and adding them up, the potential term drops out
and we end up with
\begin{equation}
J(1/\mu_a+1/\mu_p)=  \mean{\dot x_a}/\mu_a+  \mean{\dot x_p}/\mu_p=-\fe,
\end{equation}
because the averages of $\zeta_a$, $\zeta_p$, and $n$ are zero. Thus,
the current $J$ is always in the same direction as the external force,
such that work cannot be extracted.  Notably, for $\fe=0$, there is no
persistent current $J$. This result is fairly remarkable, since, according to
Pierre Curie's principle~\cite{curi94,seki10}, the asymmetry of the
potential and the nonequilibrium driving would generally be
sufficient conditions for the emergence of a persistent current.

This result can be generalised to a setting with $N$ interacting
active Brownian particles and one passive obstacle described by the
Langevin equations~\eqref{eq:langevin_cont_a}
and~\eqref{eq:langevin_cont_p} in two or three dimensions with an
isotropic mobility tensor $\vec{\mu}_a=\mu_a\vec{1}$.  The resulting
mean velocities of the active particles $J_a\equiv \mean{\dot x_a^i}$
and of the obstacle $J_p\equiv \mean{\dot x_p}$ satisfy
$NJ_a/\mu_a+J_p/\mu_p=-\fe$. If $V(\vec{r})$ is an exclusion potential
that stretches over the whole cross section of the channel or box,
such that the active particles cannot overtake the obstacle, we
have again $J=J_p=J_a$, prohibiting a positive output power.

Nonetheless, a nonzero current $J_p$ at zero external force, and thus
positive extracted power under a sufficiently small counterforce, is
achievable in several ways. First, one can choose the potential
$V(\vec{r})$ in a way that active particles can pass by or through the
obstacle, such that the currents $J_a$ and $J_p$ are no longer
constrained to be equal. Second, one can add in
Eqs.~\eqref{eq:langevin_p} and~\eqref{eq:langevin_a} an external
potential that depends explicitly on the absolute coordinates of the
particles, thus breaking the translational invariance of the system as
a whole. For instance, a periodic potential with well-separated minima
mimics the discrete lattice analysed above, showing that the lack of a
continuous translational invariance is ultimately the reason why
discrete models evade the no-go theorem. Third, one can introduce an
anisotropy in the rotational motion or a coupling or feedback between
the rotational and translational motion. Notably, this possibility
easily allows for a lossless conversion of the active power into
extracted power by fully polarising the active particles and tightly
coupling them to the obstacle.  Fourth, an anisotropic mobility tensor
$\vec{\mu}_a$, for example, due to a non-negligible $\mu_\ch$, can also
lead to a nonvanishing current against the external force.

In the following, we focus on the first possibility and consider a hard-core
interaction between the active particles and the obstacle that does not
cover the whole channel, such that active particles can pass by the
obstacle.

\section{Single active particle in continuous space}
\label{sec:single}
\subsection{General formalism}
In preparation for the many-particle case, we now study the extraction
of work in a two-dimensional setting with a single active particle
that interacts with an obstacle. Because of the periodic boundary
conditions, this setting is equivalent to an active particle
interacting with a periodic array of obstacles. For the case of
spatially fixed obstacles, experimental~\cite{volp11} and
theoretical~\cite{ghos13,jaku19} work has revealed a rich
dynamics.  

As a general consideration, we notice that, for an efficient extraction
of work, the size $L$ of the obstacle must be smaller than or
at most comparable to the persistence length $\ell\equiv u\ac/D_r$ of
the active particle. Otherwise, unless the interaction with the
obstacle affects the orientation of the active particles (not true here
with our chosen potential interaction), the active particle
behaves just like a passive Brownian particle in its interaction with
the obstacle, which cannot produce any current. Likewise, the
box length, i.e., the distance between repeated instances of the
obstacle, should not exceed the persistence length. In
reduced units, where the length scale of the obstacle and
$u\ac$ are kept fixed, the regime of high persistence corresponds to
small $D_r$, which we focus on in the following. In analogy to
the dependence of the one-dimensional system on the switching rate
$\gamma$, we expect both the power and the efficiency of the work extraction
to decrease with increasing $D_r$.

The timescale separation that ensues for small $D_r$ facilitates the
computation of the relevant currents. In two dimensions, we first keep
the vector $\vec{n}=(\cos\theta,\sin\theta)^T$ fixed and determine the
mean velocities of the active particle and the obstacle as a function
of the angle $\theta$.  Next, we account for the slow, autonomous
rotational diffusion of the active particle by averaging these
currents over $\theta$ with a uniform distribution.

We consider again the relative coordinate
$\vec{r}\equiv\vec{r}_a-\vec{r}_p$, for which the Langevin equation
follows from Eqs.~\eqref{eq:langevin_cont_p} and
\eqref{eq:langevin_cont_a} as
\begin{equation}
  \label{eq:langevin_rel}
  \dot{\vec r}=\vec{v}^0(\theta)-(\mu_a\vec{1}+\mu_p\vec{e}_x\otimes\vec{e}_x)\vec\nabla V(\vec{r})+\vec{\zeta}_a-\zeta_p\vec{e}_x.
\end{equation}
It is solved with the same periodic boundary conditions as for the absolute
coordinates. The drift terms of the active particle and the obstacle are combined to
\begin{equation}
 \vec{v}^0(\theta)\equiv u\ac\vec{n}+\mu_p\fe\vec{e}_x.
 \label{eq:v0def}
\end{equation}
The steady-state solution of the Langevin equation~\eqref{eq:langevin_rel} leads to a mean
velocity 
\begin{equation}
\vec{v}(\theta)\equiv\mean{\dot{\vec{r}}}_\theta=\vec{v}^0(\theta)-(\mu_a\vec{1}+\mu_p\vec{e}_x\otimes\vec{e}_x)\mean{\vec\nabla V(\vec{r})}_\theta,  
\end{equation}
where the
index $\theta$ indicates the ensemble average of trajectories with
fixed $\vec{n}$. 
This relation allows one to express the components of the average interaction
force between the particles in terms of the components of their
relative velocity:
\begin{equation}
  \label{eq:Vcomponents}
  \mean{\partial_x V}_\theta=\frac{v_x^0(\theta)-v_x(\theta)}{\mu_a+\mu_p},\quad  \mean{\partial_y V}_\theta=\frac{v_y^0(\theta)-v_y(\theta)}{\mu_a}.
\end{equation}
This average interaction force then yields expressions for the
average absolute velocities 
\begin{align}
  \mean{\dot{\vec{r}}_{a}}_\theta&=u\ac\vec{n}-\mu_a\mean{\vec\nabla
    V(\vec{r})}_\theta,\\
 \mean{\dot{\vec{r}}_{p}}_\theta&=\mu_p[-\fe+\mean{\partial_x
    V(\vec{r})}_\theta]\,\vec{e}_x
\end{align}
of the
individual particles. Averaging with a uniform distribution over
$\theta$ then leads to the mean velocity of the obstacle~\eqref{eq:Jcont}:
\begin{align}
  J&=\frac{1}{2\pi}\int\mrd\theta\,\mean{\dot
       x_p}_\theta\nonumber\\
  &=-\frac{\mu_a\mu_p}{\mu_a+\mu_p}\fe-\frac{\mu_p}{\mu_a+\mu_p}\frac{1}{2\pi}\int\mrd\theta\,v_x(\theta)
\label{eq:J_vrel}
\end{align}
and the extracted power $\Po=\fe J$.
The active power~\eqref{eq:Pac_cont} is given by
\begin{align}
  \Pa&=\frac{\fa}{2\pi}\int\mrd\theta\mean{\vec{n}\cdot\dot{\vec{r}}_a}_\theta\nonumber\\
&=\frac{1}{2}\frac{\mu_a\mu_p}{\mu_a+\mu_p}\fa^2+\frac{\mu_a\fa}{\mu_a+\mu_p}\frac{1}{2\pi}\int\mrd\theta\,\cos\theta\,v_x(\theta)\nonumber\\
&\qquad+\frac{\fa}{2\pi}\int\mrd\theta\,\sin\theta\,v_y(\theta),
\label{eq:Pac_vrel}
\end{align}
which is used as a reference for the active efficiency $\eta=\Po/\Pa$.
Crucially, the geometric shape of the interaction potential enters
into these expressions for the conversion of power only via the two
functions $v_{x,y}(\theta)$ for the relative velocity determined by
the Langevin equation~\eqref{eq:langevin_rel}.

\subsection{Idealised velocity filter}

With the above results at hand, we can now discuss possible shapes of
the function $v_{x,y}(\theta)$ to compare different mechanisms that
extract work through the interaction of the translational degrees of
freedom of the active particles and the obstacle.  To generate a large positive current
$J$, the integral of $v_x(\theta)$ in
Eq.~\eqref{eq:J_vrel} should be negative with a large absolute
value. Without any interaction, the relative velocity in the $x$-direction
is given by $v_x^0(\theta)=u\ac\cos\theta+\mu_p\fe$, leading
consistently to $J=-\mu_p\fe$. Broadly speaking, a well-designed
mechanism for the extraction of work should have two crucial
properties: On the one hand, when $\theta$ is such that
$v_x^0(\theta)>0$, the activity of the active particle is harnessed,
for example by trapping it in some notch of the obstacle and thereby reducing
the relative velocity to $\vec{v}(\theta)=0$. On the other hand, when
$v_x^0(\theta)<0$, the active particle should interact with the
obstacle as little as possible. If they do not interact at all, the
resulting relative velocity remains
$\vec{v}(\theta)=\vec{v}^0(\theta)$.

Without yet considering realisations of the interaction
potential that yield these properties, we can discuss the effect
of an idealised velocity filter that is accordingly modelled by
$\vec{v}(\theta)=\vec{v}^0(\theta)\chi(\theta)$. The function
$\chi(\theta)$ is set to one for $\theta_c<\theta<2\pi-\theta_c$,
when the active particle is free, and zero otherwise, when the
particle is trapped. The critical angle for which $v_x(\theta)=0$ is
given by $\theta_c\equiv\arccos(-\mu_p\fe/u\ac)$. We assume that the
external force is not exceedingly large, such that
$|\mu_p\fe|\leq u\ac$ still holds---otherwise the active particle
would be either trapped or free independently of $\theta$,
prohibiting a positive output power. Plugging the model function for
$v_x(\theta)$ into Eq.~\eqref{eq:J_vrel} yields the resulting current
$J$. It can conveniently be written as
\begin{equation}
  J=\mu_p(\fint-\fe)
  \label{eq:J_filter}
\end{equation}
with the average interaction force exerted by the active particle 
\begin{equation}
  \fint=\frac{u\ac}{\mu_a+\mu_p}\frac{1}{\pi}\left[\sqrt{1-z^2}-z\arccos(z)\right].
  \label{eq:fint_filter}
\end{equation}
The dimensionless parameter 
\begin{equation}
 z\equiv-\mu_p\fe/u\ac 
 \label{eq:zparameter}
\end{equation}
compares the velocity of the free obstacle to the active
speed. The active power~\eqref{eq:Pac_vrel} follows as
\begin{equation}
  P\ac=\mu_a\fa^2+\frac{\mu_a\fa^2}{2\pi}\frac{2\mu_a+\mu_p}{\mu_a+\mu_p}\left[z\sqrt{1-z^2}-\arccos(z)\right].
  \label{eq:Pac_filter}
\end{equation}
From Eqs.~\eqref{eq:J_filter} and~\eqref{eq:Pac_filter} we finally obtain
the expressions for the extracted power and active efficiency of the
idealised velocity filter, based only on general geometric arguments. 

\subsection{Design principles}
\label{sec:design}
With the idealised velocity filter as a benchmark, we now consider
specific realisations of the interaction potential. The optimisation
of the power and efficiency amounts to finding good designs for
isothermal engines driven by active matter. This task is related to,
but quite distinct from, the work of Ref.~\cite{van05}, which studies
design principles for ratchets driven by passive particles at two
different temperatures.

\begin{figure*}
  \centering
  \includegraphics[width=0.35\linewidth]{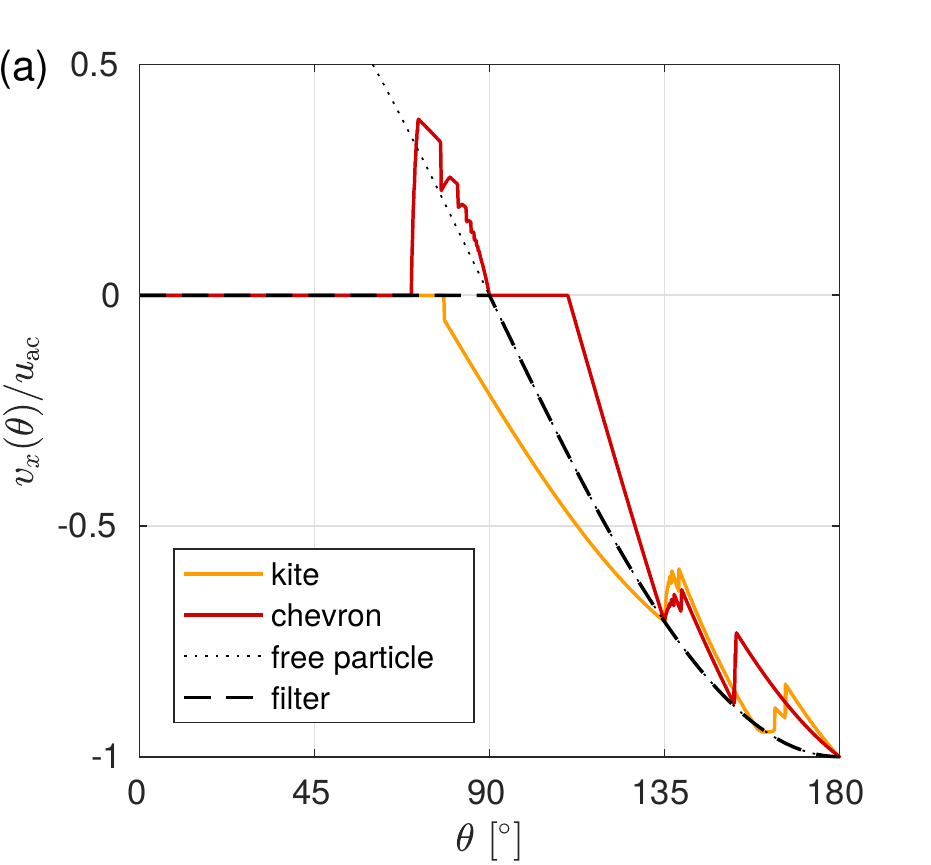}
  \includegraphics[width=0.31\linewidth]{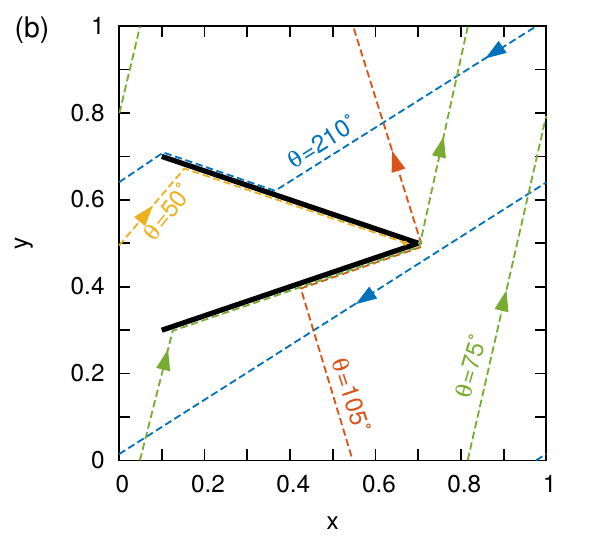}
  \includegraphics[width=0.31\linewidth]{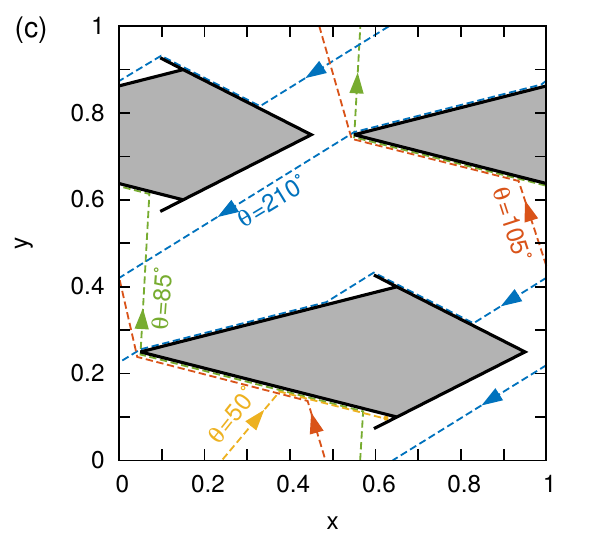}
  \caption{Angular dependence of the relative velocity of the active particle
    and the obstacle. (a) shows the velocity in the $x$
    direction for the chevron particle shown in (b) (red line) and for the kite-like
    particle shown in (c) (orange line) along with the curves for the
    ideal velocity filter (black dashed line) and the completely interaction-free particle (black dotted). For angles beyond $180^\circ$, these
    curves extend symmetrically. (b) and (c) show trajectories
    of the relative coordinate for selected angles $\theta$ to the
    $x$ axis. Except for $\theta=50^\circ$, where particles get
    trapped, we show only the periodic part of trajectories, discarding the
    initial, transient dynamics. Parameters are $\mu_p/\mu_a=0.1$ and $\fe=0$ throughout. }
  \label{fig:phi_v}
\end{figure*}

In order to keep the setting simple, we focus on hard-core
interactions and set the noise terms in the Langevin equation
\eqref{eq:langevin_rel} to zero, pertaining to a regime where the
timescale $L/|\vec{v}^0|$ of the drift process is fast compared to the
diffusion on the longer timescale $L^2/D_{a,p}$. Ensemble averages of
the type $\mean{\cdot}_\theta$ then reduce to an average over a single
periodic trajectory.  These trajectories can be calculated in a simple
numerical scheme, where the hard-core interaction is realised through
constraint forces, as detailed in Appendix~\ref{sec:hardcore}.  For
a zero external force, as explored in Fig.~\ref{fig:phi_v}, the
idealised velocity filter leads to $v_x(\theta)=u\ac\cos\theta$
for $90^\circ<\theta<270^\circ$ and $v_x(\theta)=0$ otherwise.  In
Fig.~\ref{fig:phi_v}(a), we compare this function to the numerical
results for $v_x(\theta)$ for two selected geometries of the obstacle
\footnote{Without noise, the dynamics of the relative coordinate
  $\vec{r}$ can be nonergodic for some angles $\theta$. In these cases
  we have sampled the initial value of $\vec{r}$ from the periodic
  trajectories for adjacent slightly higher and lower values of
  $\theta$, mimicking the effect of a small nonzero rotational
  diffusion coefficient.}.

A shape of obstacles that is often used to illustrate nonequilibrium
aspects of active matter is a simple \textsf{V}~shape or ``chevron''
\cite{gala07,wan08,tail09,ange10,kais12,kais13,kais14,sten16}. We
model this type of obstacle as two straight lines with a fixed opening
angle and a hard-core exclusion for the active
particles. Fig.~\ref{fig:phi_v}(b) shows this shape along with selected
trajectories of the relative coordinate~$\vec{r}$. The symmetry of
this setting allows us to restrict the discussion to angles
$0\leq\theta\leq 180^\circ$.  The chevron of the chosen geometry is
indeed capable of entrapping the active particles for positive
relative velocity $v_x^0(\theta)$ and letting it pass
otherwise. Nonetheless, the resulting function $v_x(\theta)$ differs
from the one for the idealised velocity filter in two obvious
ways. First, for angles below but close to $90^\circ$ (e.g.,
$\theta=75^\circ$), the active particle cannot be trapped; instead it
repeatedly slides along the outer side of the arms of the chevron. Accordingly, for the chosen
geometry, the function $v_x(\theta)$ is positive for
$70^\circ\lesssim\theta<90^\circ$.  Second, for angles larger than
$90^\circ$, the interaction between the active particle and the
obstacle reduces the absolute value of their relative velocity.
Notably, for $90^\circ<\theta\lesssim 110^\circ$, the repeated
interaction due to the periodic boundary conditions leads to vanishing
$v_x(\theta)$. In total, the function $v_x(\theta)$ for the
chevron-shaped obstacle is for all angles $\theta$ larger than
or equal to the one for the idealised velocity filter.  Accordingly,
the resulting current $J$ in Eq.~\eqref{eq:J_vrel} becomes reduced
compared to Eq.~\eqref{eq:J_filter}, for the chosen geometry and
parameters of Fig.~\ref{fig:phi_v} to approximately $72\%$ of the
filter value.

In principle, the output current for a chevronlike particle can be
maximised using a delicate limiting procedure. First, the opening
angle of the arms of the chevrons must be decreased to almost zero,
such that the active particle can be trapped for all angles
$\theta<90^\circ$. Second, the overall size of the chevron must be
decreased, such that the interaction between the active particle and the
obstacle for all other angles is decreased. In this limit, the
function $v_x(\theta)$ and the current $J$ approach the values
for the ideal velocity filter. However, the small size of the chevron
and its opening lead to further limitations. When the condition
$L^2/D_{a,p}\gg L/|\vec v_0|$ is no longer met, translational noise
becomes relevant, such that the active particle can be trapped only
transiently. Indeed, the small cross section of the obstacle
increases the time until the particle is trapped again, and this time
may even be comparable to the timescale set by the rotational
diffusion. It is therefore essential to first let the observation time
tend to infinity, then let the thermal noise and the rotational
diffusion coefficient tend to zero and at the very last let the size
and the opening angle of the chevron vanish. It should be noted that
in this limit, increasing the number of obstacles per unit area does
not increase the extracted power: All instances of the active particle
with $v_x^0(\theta)>0$ are ultimately trapped even in a scarce array
of obstacles, whereas it is essential that all instances with
$v_x^0(\theta)<0$ interact as little as possible with the obstacles.

Given these observations, one may be tempted to conclude that the
idealised velocity filter provides a general upper bound on the
current $J$ that can be approached only in extreme limiting
cases. Nonetheless, for more sophisticated shapes and arrangements of
the obstacle and its repeated instances, it is possible to exceed this
apparent bound. In Fig.~\ref{fig:phi_v}(c), we show a kitelike shape
with hooks at the upper and lower vertices. This shape is repeated
periodically along a square lattice that is diagonal to the direction
of motion $\vec{e}_x$ of the obstacle. We impose the
constraint that the distances to all the repeated instances of the
obstacle are kept fixed over time, such that it is still
sufficient to describe the position of the ensuing array of obstacles
with a single variable $x_p$.

The hooks at the upper and lower vertices of the kite-shaped particle
take over the role of the chevrons in trapping the active particle for
angles in the region around $\theta=0$. For the chosen geometry, this
trapping ensues for angles $|\theta|\lesssim 80^\circ$. Crucially, for angles
somewhat above this threshold, the elongated rear shape of the kites
and the pattern in which they are arranged force the coordinate
$\vec{r}$ on a trajectory whose general direction is $(-1,1)$, thus
reversing the sign of $v_x(\theta)$ compared to $v_x^0(\theta)$. As
visible in Fig.~\ref{fig:phi_v}(a), this effect persists for all angles
up to approximately $135^\circ$, leading to negative velocities
$v_x(\theta)$ below the curve for the idealised velocity filter. For
even larger values of $\theta$, an interaction between the active
particle and the obstacle leading to $v_x(\theta)>v_x^0(\theta)$ cannot be avoided. Nonetheless,
when averaging over all $\theta$, a positive effect prevails. The
width and length of the kite shown in Fig.~\ref{fig:phi_v}(c) have been
optimised to yield a current that is approximately $5\%$ larger than
that of the idealised velocity filter (with fixed $\mu_p/\mu_a=0.1$
and $\fe=0$). The overall proximity between
the functions $v_x(\theta)$ for the kite-shaped particle and the
velocity filter justifies the role of the latter as an analytically
tractable model for the thermodynamics of a well-designed work
extractor.

\begin{figure}
  \centering
  \includegraphics[width=\linewidth]{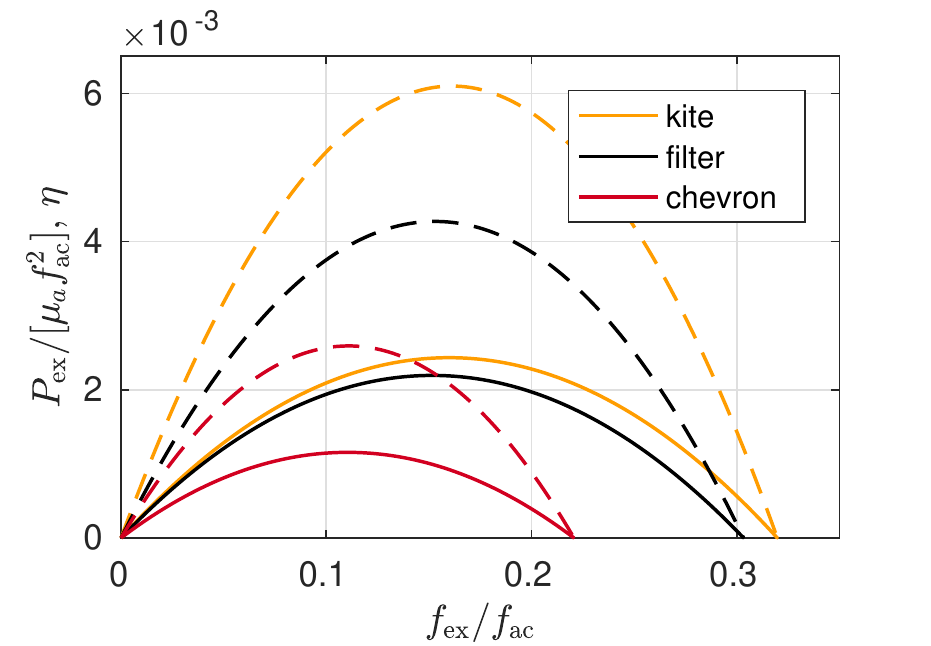}
  \caption{Output power (solid curves) and active efficiency (dashed curves) as a
    function of the external force $\fe$ for the velocity filter
    (black curves)
    chevron (red curves) and kite (orange curves) from Fig.~\ref{fig:phi_v} for a
    single active particle. The ratio of mobilities is kept fixed at
    $\mu_p/\mu_a=0.1$.}
  \label{fig:f_P}
\end{figure}

Next, we explore the dependence of the current $J$ on the external
force $\fe$. For this purpose, we make use of the fact that a change
of $\fe$ in Eq.~\eqref{eq:v0def} has the same effect as a
change of the angle $\theta$ and the speed $u\ac$. Making explicit the
dependence of $\vec{v}$ on $\theta$, $\fe$, the potential, and the
noise, this correspondence can be expressed as
\begin{equation}
  \vec{v}(\theta,\fe,V,\zeta_a,\vec{\zeta}_p)=\alpha\vec{v}(\tilde
  \theta,0,V/\alpha,\zeta_a/\alpha,\vec{\zeta}_p/\alpha),
\label{eq:phitrafo1}
\end{equation}
with
\begin{equation}
  \tan\tilde\theta=\frac{\sin\theta}{\cos\theta-z},\
  \alpha=\sqrt{\sin^2\theta+(\cos\theta-z)^2}
\label{eq:phitrafo2}
\end{equation}
and the scaled external force $z$ as above.  In particular, for a
hard-core interaction and in the absence of noise, as discussed above,
the knowledge of the function $\vec{v}(\theta)$ at zero external force
is sufficient to calculate the integrals in Eqs.~\eqref{eq:J_vrel}
and~\eqref{eq:Pac_vrel} for arbitrary $\fe$.  The loading curves in
Fig.~\ref{fig:f_P} show the results for the extracted power and the
active efficiency for the chevron and kitelike shapes from
Fig.~\ref{fig:phi_v}, which are compared to the analytical expressions
for the idealised velocity filter
derived from Eqs.~\eqref{eq:J_filter} and \eqref{eq:Pac_filter}.  

We observe that the extracted power is rather small. Taking the active
power of a free particle $\mu_a\fa^2$ as a reference, the scaled
extracted power $\Po/\mu_a\fa^2$ does not exceed $0.0025$ for
$\mu_p/\mu_a=0.1$ as chosen in Fig.~\ref{fig:f_P}. For the velocity
filter, a global maximisation yields the bound
$\Po\lesssim 0.0089\,\mu_a\fa^2$, which is reached for
$\mu_p/\mu_a\simeq 1.48$ and $\fe/\fa\simeq 0.094$. The values for the
active efficiency are larger than $\Po/\mu_a\fa^2$, because the
interaction between the particles reduces the active power compared to
the free active particle.  The superiority of the kite-shaped work
extractor persists for all external forces, producing a larger
current, power, and efficiency than the velocity filter.

\section{Many active particles}
\label{sec:many}
\subsection{Mean-field theory}
\label{sec:MF}

Building on the results for the single active particle, we next study the extraction of work in a setting with a large
number $N$ of active particles. We focus on the dilute limit, where
the size of the active particles is assumed to be sufficiently small
compared to the typical interparticle distances. The direct
interactions among the active particles can then be
neglected. Nonetheless, the small active particles still interact with
a large obstacle. Naively, one might expect that this
interaction is simply additive in the number of active
particles. However, we have to take into account that each active
particle affects the motion of the obstacle, which has, in turn, some
effect on its interaction with all other active particles.

Focusing on noninteracting active particles, the number density of
obstacles plays only a subordinate role. Since an obstacle can, in
principle, trap arbitrarily many active particles, it is irrelevant
whether a single obstacle extracts power from all active particles
together or whether several obstacles each extract only a fraction of
the power. The only requirement, as before for the single-particle
case, is that the distance between obstacles does not exceed the
persistence length of the active particles.

In order to derive the key quantities, we use a mean-field approach,
focusing first on the interaction between a representative active
particle and the obstacle, where the influence from all other
active particles is subsumed with the external force. \textit{A
  posteriori}, this influence is determined self-consistently.

We start with some general considerations for the dynamics of the
obstacle interacting with a background of many small active
particles and producing work against a counterforce. First, we notice
that the velocity of the obstacle cannot persistently exceed
the velocity $u\ac$ of the individual active particles; otherwise all
interactions would be directed against the direction of motion of the
obstacle. Thus, one can only hope to increase the extracted
power with the number of active particles by simultaneously increasing
the external force.

Second, the obstacle and the many active particles currently pushing
it are in a close contact, such that they may be regarded as a
single complex. Since we neglect hydrodynamic interactions, the
friction coefficients of the objects forming such a complex are
additive. As the number of active particles in this complex increases
linearly with $N$, we can assign to the interacting obstacle an
effective friction coefficient, or inverse mobility, that scales also
linearly in $N$. This scaling later turns out to be self
consistent in the mean-field analysis. Since the forces acting on the
complex of obstacle and active particles also scale linearly in $N$,
the resulting average velocity $J$ can still remain nonzero.

Third, we expect that fluctuations in the dynamics of the obstacle
vanish in the limit of many active particles. Such fluctuations have
two sources: The thermal noise acting on the complex of the obstacle
and trapped active particles scales according to the effective
mobility like $1/\sqrt{N}$. The other contribution stems from
fluctuations of the force exerted by all the active particles. Being
the sum of $N$ independent random variables, the fluctuations in the
resulting force scale like $\sqrt{N}$. Multiplication by the effective
mobility shows that the impact of these fluctuations on the velocity
of the obstacle vanishes also like $1/\sqrt{N}$.

As a result of the above considerations, we can replace the Langevin
equation \eqref{eq:langevin_cont_p} for the obstacle in the
mean-field limit by a simple motion $\dot{\vec r}_p=J\vec{e}_x$ with a
constant, yet-to-be-determined velocity $J$. The form of the Langevin
equation \eqref{eq:langevin_cont_a} for a representative active
particle, with the interaction term $U$ set to zero, is unaffected by
the presence of the other active particles. The many-body dynamics
then reduces to an effective two-body problem analogous to the one in
Sec.~\ref{sec:single}. The Langevin equation for the relative
coordinate $\vec{r}$ between the representative active particle and
the obstacle follows as
\begin{equation}
 \dot{\vec{r}}=\tilde{\vec{v}}^0(\theta)-\mu_a\vec\nabla V(\vec{r})+\vec{\zeta}_a.
 \label{eq:mf_langevin}
\end{equation}
It has the same form as Eq.~\eqref{eq:langevin_rel} for the
single active particle, but with $\mu_p$ and, thus, $\zeta_p$ set to
zero and the drift term redefined as
\begin{equation}
  \tilde{\vec{v}}^0(\theta,J)\equiv u\ac\vec{n}-J\vec{e}_x.
\end{equation}
The solution of the Langevin equation~\eqref{eq:mf_langevin} leads to
the stationary relative velocity $\vec{v}(J)=\mean{\dot{\vec{r}}}$,
where we make the dependence on $J$ explicit. 

We stress that our mean-field approach is not limited to persistent
active particles with a timescale separation between translational and
rotational motion, as considered in Sec.~\ref{sec:single}.  In
general, Eq.~\eqref{eq:mf_langevin} is to be solved with rotational
diffusion in the angle $\theta$, which is thereby averaged over in the
computation of $\vec{v}(J)$. In case we do have very persistent active
particles, we can use the same strategies as before for the
single-particle case, starting with a model function or explicit
results for $\vec{v}(\theta)$, obtaining the dependence on $J$ through
the transformation \eqref{eq:phitrafo1} with $z=J/u\ac$, and then
integrating out $\theta$.

The interaction force exerted by the
representative active particle in the $x$ direction on the obstacle
follows from Eq.~\eqref{eq:mf_langevin} as
\begin{equation}
  \fint(J)\equiv\mean{\partial_x
    V(\vec{r})}=-\left[J+v_x(J)\right]/\mu_a.
  \label{eq:mf_fint}
\end{equation}
In the mean-field solution, this force is exerted by each of the
active particles, which together yield the total force acting on the
bare obstacle. Consistency with the generally valid
relation~\eqref{eq:Jcont} therefore requires
\begin{equation}
  J=\mu_p[-\fe+N\fint(J)],
  \label{eq:selfconsistence}
\end{equation}
which finally relates $J$ to the corresponding external force $\fe$
and yields the extracted power 
\begin{equation}
\Po=\fe J=[N\fint(J)-J/\mu_p]\,J.  
\label{eq:Pomf}
\end{equation}
Moreover, consistently with what
we have assumed before, the effective mobility of the obstacle
in contact with the active particles scales like
\begin{equation}
 \mu_{p,\mathrm{eff}}\equiv-\frac{dJ}{d\fe}=\left(\frac{1}{\mu_p}-N\fint'(J)\right)^{-1}\sim 1/N .
\end{equation}

On the other hand, the total active power, as defined in Eq.~\eqref{eq:Pac_cont}, is
given in the mean-field limit from the $N$ independent contributions
of all active particles as
\begin{equation}
  \Pa=N\fa\mean{\vec{n}\cdot[J\vec{e}_x+\dot{\vec{r}}]}=N\fa\mean{\vec{n}\cdot\dot{\vec{r}}},
\label{eq:Pamf}
\end{equation}
where the averages are computed from the Langevin equation
\eqref{eq:mf_langevin} and run over all angles $\theta$, such that $\mean{\vec{n}\cdot\vec{e}_x}=0$.

\begin{table}
  \centering
  \begin{tabular}{c||c|c}
    $N$ & $1$ & Many \\\hline
$P_\mathrm{ex,max}/(N\mu_a\fa^2)$ & $0.0089$ & $0.058$ \\
$\fe^*/(N\fa)$ & 0.094 & 0.15\\
$(\mu_p/\mu_a)^*$ & 1.5 & $>\mathcal{O}(1/N)$ \\
$\eta^*$ & $1.5\%$ & $7.7\%$\\
$\eta_\mathrm{max}$ & $1.5\%$ & $8.0\%$
  \end{tabular}
  \caption{Thermodynamic characterisation of the extraction of
    work for the idealised velocity filter in
    a setting with a single active particle ($N=1$) and in the limit
    of many active particles. Listed are the maximal extracted power per
    active particle (along with the maximising parameters $\fe$ and
    $\mu_p$), the active efficiency at maximum power $\eta^*$, and the maximal
    active efficiency $\eta_\mathrm{max}$. Note that, in the mean-field
    theory, the maximum power is independent of the mobilities
    $\mu_{p,a}$, as long as their ratio is well above $1/N$. }
  \label{tab:results}
\end{table}

\begin{figure}
  \centering
  \includegraphics[width=\linewidth]{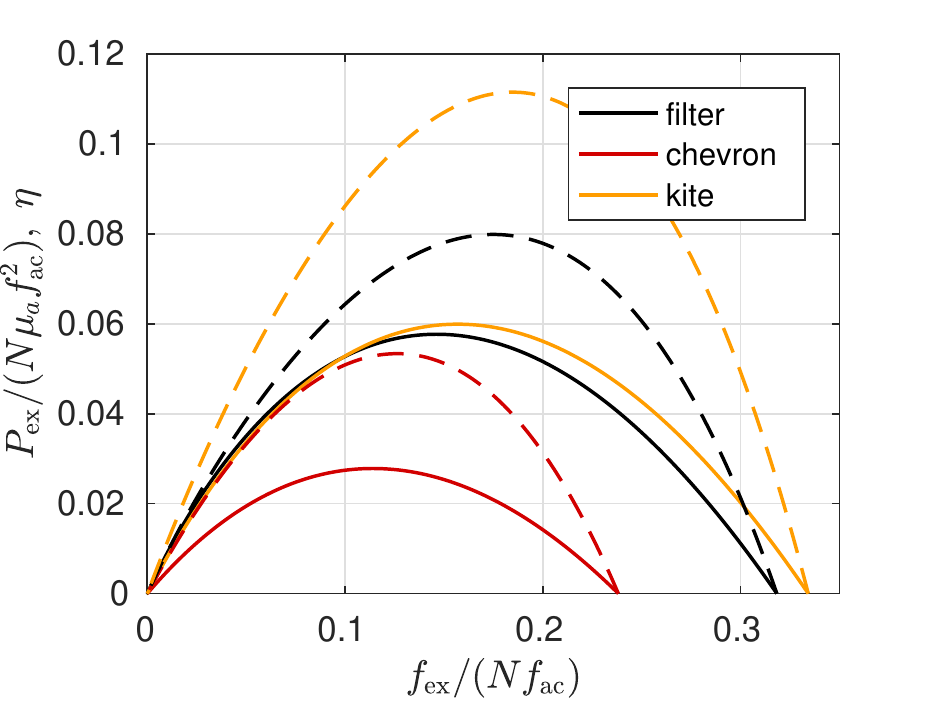}
  \caption{Output power (solid curves) and active efficiency (dashed curves) as a
    function of the external force $\fe$ for the velocity filter
    (black curves)
  chevron (red curves) and kite (orange curves) from Fig.~\ref{fig:phi_v} in the
  mean-field limit of many active particles with large persistence.}
  \label{fig:f_P_meanfield}
\end{figure}

For the idealised velocity filter, we focus again on persistent active
particles and calculate 
\begin{equation}
  \vec{v}(J)=\frac{1}{2\pi}\int_{\theta_c}^{2\pi-\theta_c}\mrd\theta\,\vec{v}^0(\theta,J),
\end{equation}
where the critical angle is defined through
$v_x^0(\theta_c)=u\ac\cos\theta_c-J=0$. Carrying out the integration,
we obtain 
\begin{equation}
  \fint=\frac{u\ac}{\mu_a\pi}\left[\sqrt{1-z^2}-z\arccos(z)\right],
  \label{eq:fint_filter2}
\end{equation}
which is similar to Eq.~\eqref{eq:fint_filter} for the single-particle
case but with a redefined dimensionless parameter $z\equiv J/u\ac$. Again, we focus on $|z|\leq1$,
corresponding to the region of interest where $|J|\leq u\ac$. Rather then solving the ensuing
transcendental equation for $J$ [Eq.~\eqref{eq:selfconsistence}], we can analyse the dependence of
the extracted and active power on the external force in terms of parametric
plots defined by
\begin{subequations}
\begin{align}
  \fe(z)&=\frac{N\fa}{\pi}\left[\sqrt{1-z^2}-z\arccos(z)-\frac{\pi\mu_a}{N\mu_p}z\right],\\
  \Po(z)&=u\ac z\,\fe(z),\\
  \Pa(z)&=\frac{N\mu_a\fa^2}{\pi}\left[\pi+z\sqrt{1-z^2}-\arccos(z)\right];
\end{align}
\label{eq:parametric}
\end{subequations}
see Fig.~\ref{fig:f_P_meanfield}. 

The only parameter that does not amount to a mere overall scaling of the
above equations is $\lambda\equiv\mu_a/(N\mu_p)$. Note that $\mu_p$ is
here the bare mobility entering through Eq.~\eqref{eq:selfconsistence}
and not the vanishing effective one. Provided that the ratio of bare
mobilities $\mu_a/\mu_p$ is not of the order of $N$, we can set
$\lambda=0$, leaving us with a parameter-free
representation. Otherwise, for a large obstacle that is much
less mobile than the active particles, an
ensuing positive value of $\lambda$ reduces $\Po(z)$ for all $z$, leading to a
smaller maximal extracted power. The extracted power is maximised for
$z^*=\cos(y^*)\simeq 0.394$, where $y^*$ is the smallest positive
solution of $2y=\tan y$. The external force corresponding to $z^*$ is
given by $\fe(z^*)=(N\fa/2\pi)\sin y^*\simeq 0.146\,N\fa$. The maximal
extracted power itself is
$\Po(z^*)=(N\mu_a\fa^2/2\pi)z^*\sin y^*\simeq 0.0577N\mu_a\fa^2$, and
the active power is $\Pa(z^*)\simeq 0.744N\mu_a\fa^2$, leading to an
active efficiency at maximum power of $\eta^*\simeq 7.74\%$. This
result is
only little below the maximal active efficiency
$\eta_\mathrm{max}\simeq 7.99\%$ that is reached for
$\fe\simeq0.175\,N\fa$ and $\lambda=0$, for which the active power is
$\Pa\simeq 0.0559N\mu_a\fa^2$.

We recall that in Sec.~\ref{sec:single} the power extracted from a
single active particle is rather small, amounting to roughly $1\%$ of
the active power expended by the active particle. Naively, one may
have expected that by using $N$ noninteracting active particles both
the extracted and the expended power increase linearly, leading to a
similarly small efficiency.  Surprisingly, however, as summarised in
Table~\ref{tab:results}, we find analytically for the idealised
velocity filter that the extractable power per active particle and the
characteristic efficiencies are consistently higher by nearly one
order of magnitude in the setting with many active particles than in
the one with a single active particle. This increase is an important
result of our paper, which could not have been anticipated \textit{a
  priori}.

The joint interaction with the obstacle mediates some kind of
cooperativity between the otherwise noninteracting active
particles. For an intuitive understanding of this behaviour, consider
the reaction of the obstacle to the detachment of a previously
trapped active particle. If there are no other active particles, the
obstacle is then surrendered completely to the external force
pulling it backwards. Such negative contributions to the extracted
power are prevented when the presence of many more trapped active
particles stabilises the forward motion of the obstacle. Beyond
active matter, the collective effects observed here are somewhat
reminiscent of the ones observed in coupled molecular
motors~\cite{juel95} and, more recently, in coupled heat
engines~\cite{vroy17} and power converters~\cite{herp18}.

As before, the idealised velocity filter serves as a benchmark for the
performance of work extractors based on suitably shaped obstacles. In
Fig.~\ref{fig:f_P_meanfield}, we compare its power and efficiency to
that of the chevron and kite-shaped particles in the mean-field
limit. For this purpose, we solve Eq.~\eqref{eq:mf_langevin} for
the two geometries shown in Figs.~\ref{fig:phi_v}(b) and~\ref{fig:phi_v}(c) and with the noise
term set to zero. These solutions yield velocity profiles similar to the ones in
Fig.~\ref{fig:phi_v}a, which can be used to compute $\vec{v}(J)$ along
with the relevant thermodynamic quantities.

For the chevron particle, the mean relative velocity $v_x(J)$ is always
larger than for the idealised velocity filter. Hence, the interaction force~\eqref{eq:mf_fint} is
below that of the velocity filter for any given current $J$. In
Eq.~\eqref{eq:selfconsistence}, this difference leads to a
smaller corresponding external force $\fe$ and, thus, a smaller extracted
power $\Po=\fe J$. In contrast, the kite-shaped particle has a somewhat higher
maximal extracted power than the velocity filter. For small external
forces though, the extracted power is somewhat smaller. This regime
corresponds to large currents $J$, where the kite-shaped particle,
unlike the idealised velocity filter, experiences strong ``headwind''
from surrounding active particles. In comparison to
Fig.~\ref{fig:f_P}, we stress that for both designs of the
obstacle the attained efficiency and power per active particle are
larger than in the case with a single active particle.

\subsection{Numerical simulations}

To test our design principles in actual many-particle settings, we now
turn to the numerical study of autonomous engines driven by a bath of
active particles. We consider a set of noninteracting active Brownian
particles in two dimensions with position dynamics given
by Eq.~\eqref{eq:langevin_cont_a}. As usual for this type of model, the
translational noise ${\boldsymbol\zeta}_{a}^i$ is assumed to have
isotropic Gaussian correlations, which amounts to neglecting the
chemical mobility compared to the thermal one
($\mu_{\rm ch}\ll\mu_{\rm th}$). We allow the angular direction
$\vec{n}^i = (\cos\theta^i, \sin\theta^i)$ to fluctuate in time
following an independent dynamics for each particle:
\begin{equation}
	\dot \theta^i = \sqrt{2 D_{r}} \xi^i ,
\end{equation}
where $D_{r}$ is the rotational diffusion coefficient. The noise
term $\xi^i$ has Gaussian statistics with zero mean and variance given
by $\langle\xi^i(t)\xi^j(t')\rangle = \delta_{ij}\delta(t-t')$. We
recall the definition of the according persistence length
$\ell =u\ac/ D_{r}$ as the typical distance
covered by a particle, in the absence of an obstacle, before changing its
orientation.

\begin{figure*}
	\centering
	\includegraphics[width=.37\columnwidth]{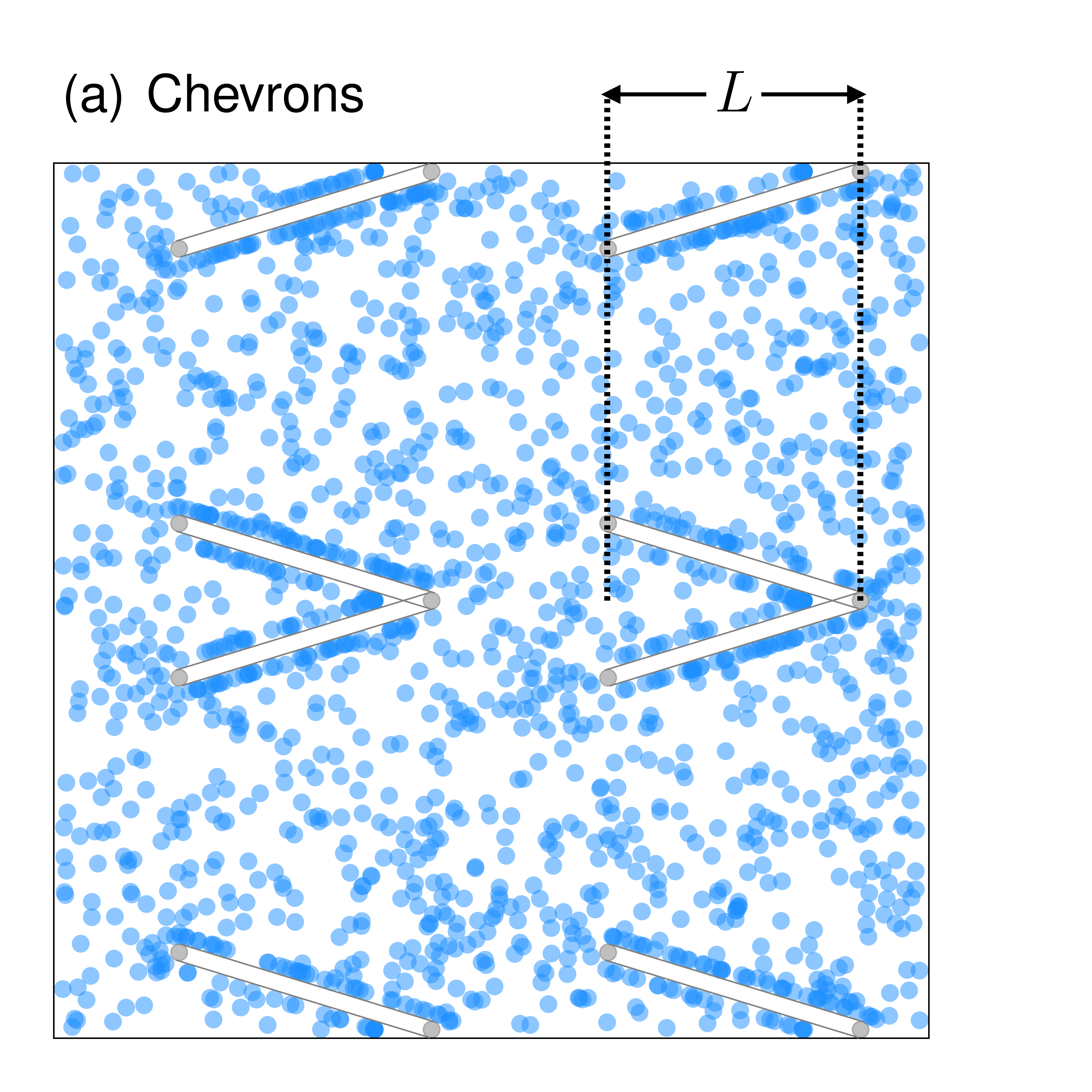}
	\hfill
	\raisebox{.35cm}{\includegraphics[width=.61\columnwidth, trim = 0 0 10cm 0, clip=true]{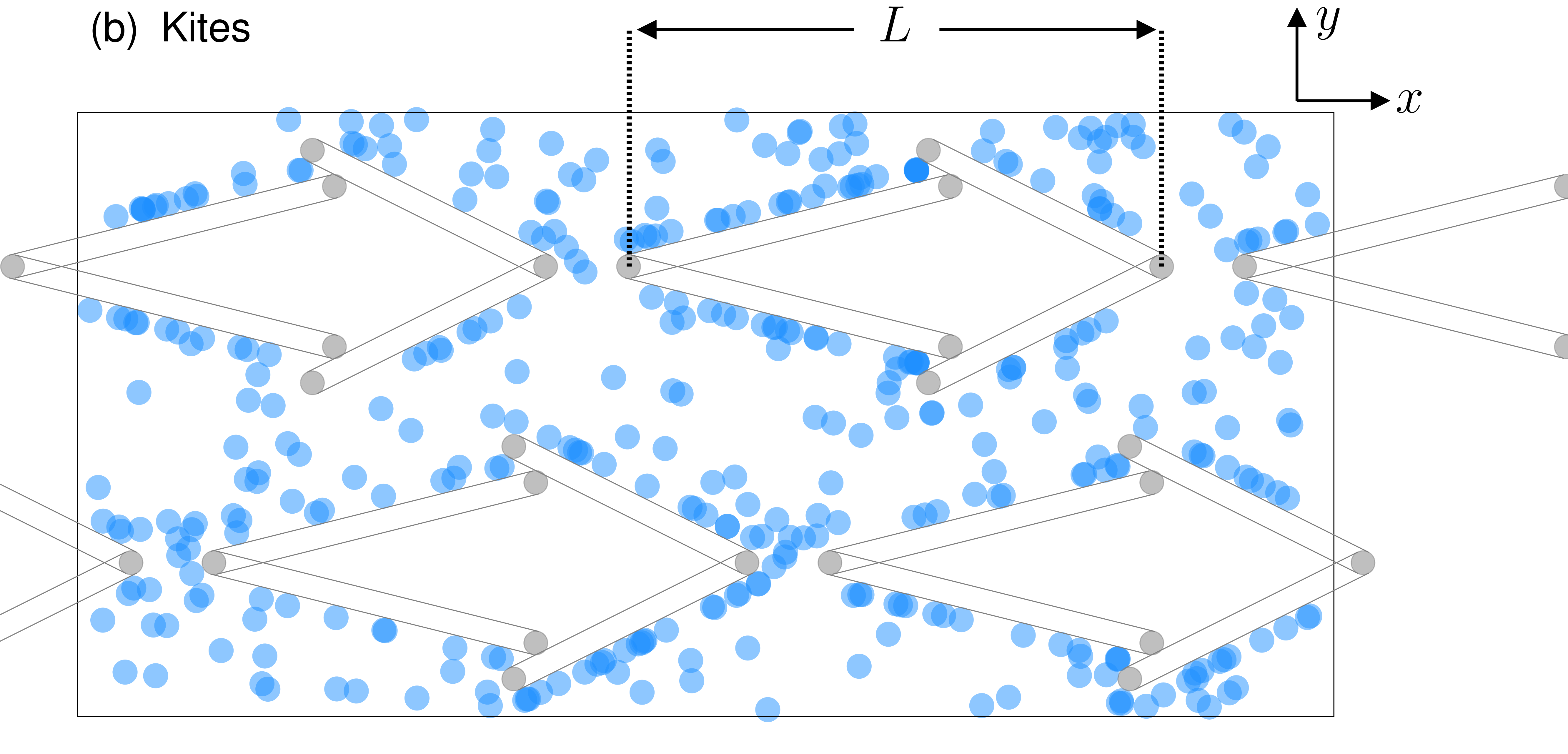}}
	\includegraphics[width=.49\columnwidth]{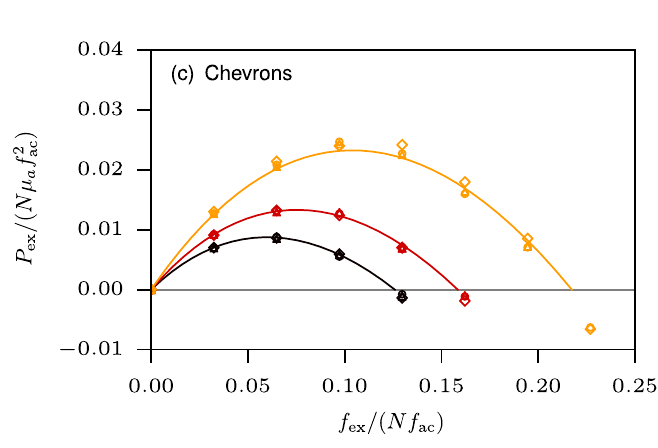}
	\hfill
	\includegraphics[width=.49\columnwidth]{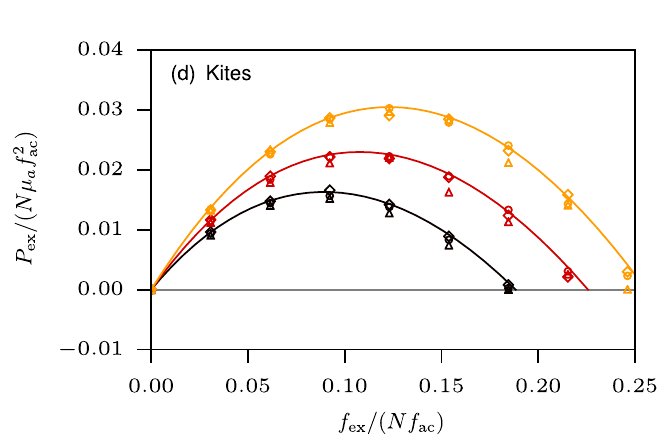}
	\includegraphics[width=.49\columnwidth]{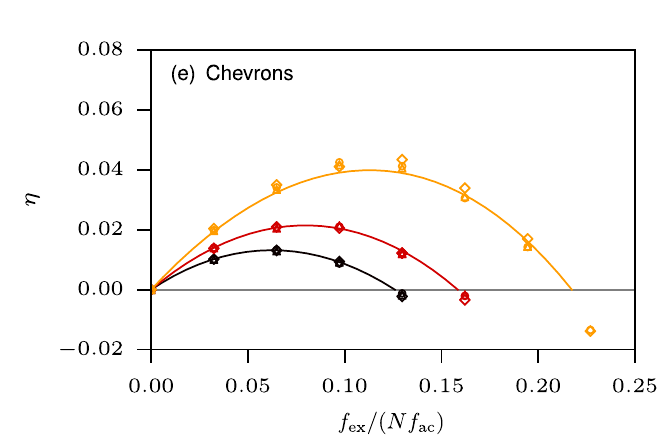}
	\hfill
	\includegraphics[width=.49\columnwidth]{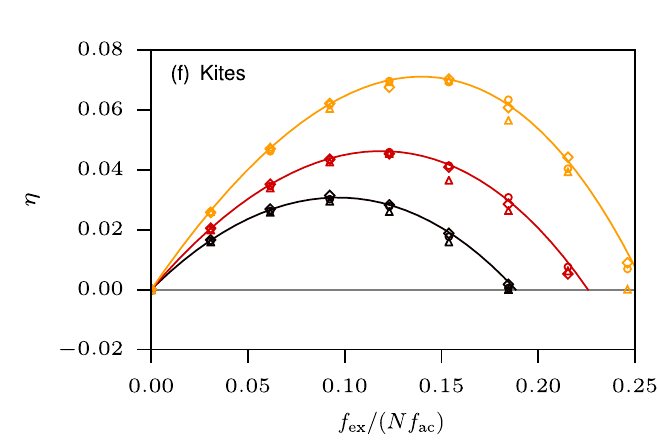}
	\caption{\label{fig:simu} Performances of autonomous engines
          in a bath of active particles. The engines are made of a
          series of asymmetric obstacles, either (a)~chevrons or
          (b)~kites, with large axis denoted by $L$. The displacement
          of all obstacles is synchronised and restricted to the $x$
          axis. Active particles, shown as blue circles, interact only
          with the obstacles. Using a biperiodic box with size
          $L_x\times L_y$ and given the shape asymmetry, the obstacles
          follow a perpetual directed motion towards $x>0$. To extract
          work, the operator applies a constant force
          $f_{\rm ex}$ towards $x<0$ on the obstacles.  The extracted power
          $P_{\rm ex}$ and the efficiency $\eta$ are, respectively,
          shown for (c, e)~chevrons and (d, f)~kites as functions of
          the applied force.  The shapes of the symbols refer to
          particle density $\rho$ accounting for the excluded obstacle
          area: $\rho=0.23$ (triangles), $0.46$ (circles), and $0.68$
          (diamonds). The colour code corresponds to persistence lengths
          $\ell=\mu_{a}f_{\rm ac}/D_{r}$ of the active particles. Chevrons:
          $\ell/L=3.3$ (black), $6.6$ (red), and $66$
          (orange). Kites: $\ell/L=2.2$ (black), $4.4$ (red),
          and $44$ (orange). Solid curves are the predictions of the
          mean-field theory, obtained from simulations for a single
          active particle.  Other parameters: $D_p=10^{-2}=D_a$
          (except for chevrons at $\ell/L=66$, where
          $D_p=1=D_a$), $f_{\rm ac}=1$, $\mu_p=1=\mu_a$, $V_0=10^2$,
          and $L_x\times L_y=52\times 52$ (chevrons) and $52\times 25$
          (kites).  }
\end{figure*}

We now model each obstacle by an assembly of soft rods which interact
repulsively with the surrounding active particles. The potential
between a particle $i$ and a rod $j$ is taken as short ranged of the
form $V(r_{ij}) = V_0 ( 1 - r_{ij}/a )^2 $ for $r_{ij}<a$, where
$r_{ij}$ is the minimal distance between the particle centre and the
points on the line segment of the rod. In practice, we use $a=1$
in what follows, so that all length scales are expressed in units of the
particle-rod interaction length. Besides, the energy scale $V_0$ is
always large compared with the ones of thermal fluctuations $k_BT$ and
the active force $a\fa$, so that the rods effectively act as
hard walls. Following the geometry introduced in
Sec.~\ref{sec:single}, we can then form two types of obstacle, either
chevrons or kites, as shown in Figs.~\ref{fig:simu}(a) and \ref{fig:simu}(b). The arrangement
of the obstacles is directly inspired by the periodic structures in
Sec.~\ref{sec:single}, namely, a simple square lattice for chevrons and
a two-lane arrangement for kites, and it is kept fixed throughout the
simulations. The displacement of all rods forming the obstacles is
synchronised and restricted to the $x$ direction with dynamics given
by~\eqref{eq:langevin_cont_p}. Finally, we use biperiodic
boundary conditions, so that the obstacles follow a perpetual directed
motion towards $x>0$ in the absence of external force
($f_{\rm ex}=0$) (See Supplemental Material\footnote{Supplemental
  Material for movies corresponding to Figs.~\ref{fig:simu}(a,b) is
available at \url{http://link.aps.org/supplemental/10.1103/PhysRevX.9.041032}}).

We measure the extracted power per active particle and the efficiency
as functions of the external force for both chevrons and kites, as
reported in Figs.~\ref{fig:simu}(c)-\ref{fig:simu}(f). At a given
value of the persistence length $\ell$, the loading curves extracted
from various numbers of active particles $N$ fall onto a master curve,
in agreement with the mean-field regime considered in
Sec.~\ref{sec:MF}. When increasing the persistence length $\ell$, the
stall force, the maximum power and efficiency as well as the
corresponding force values increase. These data corroborate that the
regime of large persistence is indeed optimal, as we assume in
Sec.~\ref{sec:single}. In practice, the orange curves corresponding in
Fig.~\ref{fig:simu} to the largest persistence coincide with the ones
for infinite persistence, namely, when $D_{r}=0$.  The translational
diffusion coefficients $D_a$ and $D_p$ have only little influence on
the loading curve, as long as the thermal energy is small compared to
the energy required for a particle to leave a trapped state.
Moreover, the peak values of power and efficiency are systematically
higher for kites compared with chevrons. This difference shows that
the kites achieve better performances not only at a large persistence,
but also for intermediate regimes. In short, these numerical results
demonstrate that the design principles we put forward indeed allow one
to delineate the optimal geometry for autonomous engines in a
fluctuating active bath.

Comparing the loading curves in
Figs.~\ref{fig:simu}(c)-\ref{fig:simu}(f) with the corresponding
analytic predictions in Fig.~\ref{fig:f_P_meanfield}, the peak values
extracted from numerical simulations turn out to be smaller. Two
reasons account for this difference. First, our simulations include
explicit fluctuations, which are neglected in the previous analytic
treatment, and which lower the maxima of the loading curves at
intermediate persistence. Second, the obstacle geometries differ
somewhat in the simulations compared with the pictures in
Fig.~\ref{fig:phi_v}. This difference is due to the finite size of
active particles and finite width of rods, in contrast with the
pointlike and linelike approximation used in Secs.~\ref{sec:design}
and~\ref{sec:MF}. While our simulations serve as a proof of principle,
further improvements of the power and efficiency may be expected for a
rigorous optimisation of the obstacles' shape and arrangement under
the constraints set by such a more realistic setting.

For a quantitative verification of the mean-field approach, we
evaluate the single-particle dynamics by numerically integrating the
Langevin equation~\eqref{eq:mf_langevin} specifically for the geometry
and diffusion coefficients used in the simulation and for a finely
discretised set of values for $J$. The force, extracted power, and
efficiency in the mean-field limit of many active particles then
follow from Eqs.~\eqref{eq:selfconsistence}, \eqref{eq:Pomf}, and
\eqref{eq:Pamf}. The loading curves resulting as parametric plots are
shown as solid curves in Figs.~\ref{fig:simu}(c)-\ref{fig:simu}(f). They agree well
with the results of the simulation, indicating that the particle
densities used in the simulation are already sufficiently large to
justify the mean-field assumptions. In particular, since the
mean-field theory works even for the lowest density used in the
simulations, one can expect that the enhancement of the power and
efficiency are observable even in a dilute realisation of a model with
interactions between active particles, before clogging effects
decrease the power again at very high densities~\cite{reic18}.

\section{Conclusions}
\label{sec:conclusions}

In this work, we have analysed the dynamics and energetics of
asymmetrically shaped passive obstacles immersed in active baths. The
interaction with active particles propels the obstacles such that they
can deliver work against a mechanical counterforce. In such a
setting, thanks to the simultaneous breaking of spatial and
time-reversal symmetries, the obstacles act as autonomous engines
driven by active matter. This type of setting is minimal for an engine
driven by an active bath; in stark contrast to classical heat engines,
it requires neither a second bath nor any cyclic manipulation of
system parameters.

In a general approach, we have identified the quantities that are
relevant for a characterisation of the thermodynamics of active
engines. An obvious quantity to consider is the extracted work,
defined as the external counterforce times the displacement of the
obstacle. For a quantification of the efficiency of an engine, this
work is to be compared to the input of energy. Yet, the total rate at
which chemical energy is supplied to maintain the active particles'
self-propulsion is hard to assess, as it typically involves many
unresolved microscopic processes. Moreover, most of this chemical
energy is typically dissipated on a microscopic scale and can,
therefore, fundamentally not be extracted by any mechanism that
operates on a mesoscopic scale. In contrast, the active work we
have considered here is a more easily assessable quantity at a
mesoscopic level, which also turns out to be more closely related to
the extracted work. It takes into account the displacement of active
particles driven by an effective active force, which can be
inferred phenomenologically from experimental data.  Here, we have
shown how the active work can be identified through coarse-graining
from a minimal, microscopic, and thermodynamically consistent model
for active particles. As a result, the commonly used active Brownian
particle model emerges in a way that allows us to disentangle
chemical aspects of entropy production from coarse-grained ones.  We
thus have formalised the concept of active force
\cite{Fily2012,Redner2013,magg15,fara15,fodo18a} and
work~\cite{fodo14,fodo16,cagn17,nemo19}, previously used in
theoretical models, from a thermodynamic perspective.  Moreover, we
have shown how the work that can be extracted on a macroscopic scale
is related to this active work. Since the former is less than the latter,
we can define the active efficiency as the ratio of these two
quantities. It is an upper bound on the ``full'' thermodynamic
efficiency defined as the ratio of extracted work to the chemical
energy expended microscopically, which is, however, typically not measurable. In
contrast, the active efficiency allows an experimenter who has access
to an active bath to quantify the performance of an engine built on
it, independently of mesoscopically irrelevant chemical details of the
particles' self-propulsion mechanism.

We have investigated the power and efficiency of work extraction from
active matter for minimal examples of engines in various
settings. Common to all of these settings is the fact that the
extracted power increases with the persistence length of active
particles. For a one-dimensional lattice model with one active
particle and the obstacle represented by a passive particle with
asymmetric interactions, we have calculated the power and efficiency
exactly. In one limiting case, the active efficiency reaches unity,
revealing that there can be no stronger universal bound on the
extracted power.

For a fairly general Langevin model in continuous space, a no-go
theorem shows that power can be extracted only when active particles
have the possibility to pass by the passive obstacle. Therefore, we have
focused on two-dimensional settings, where such a passing by
is possible even for particles with hard-core interactions.

For the case of a single active particle and a single passive
obstacle, we have considered the effect of the geometry of the
obstacle on the power and efficiency. An analytically solvable
benchmark is given by an obstacle with the idealised behaviour of a
velocity filter, trapping particles moving in one direction and
letting pass particles in the other. Simple chevron-shaped particles
cannot surpass the power and efficiency of such a filter. Nonetheless,
we have shown that, with a more complex design of obstacles, it is
possible to improve upon this benchmark by a small margin.

For obstacles immersed in a bath of many active particles that do
not interact with each other, we have calculated the power and
efficiency of the work extraction using a mean-field approach. It
reveals that at high number densities the efficiency and the power per active particle
are enhanced by one order of magnitude compared to the case of a single
active particle. Numerical simulations for the many-particle setup
validate the mean-field approach.

In this paper, our illustrations of work-extraction mechanisms have
been focused on highly idealised model systems. For instance, we have
not considered pair interactions between active particles, alignment
interactions between the active particles and the obstacle, or hydrodynamic
interactions, which would likely all be present in experimental
realisations. These idealisations have allowed us to obtain analytical
results and general design principles. Nonetheless, one may expect
that these results provide benchmarks for a more general class of
models, in which our idealisations are embedded as limiting cases, in
particular the dilute limit. 

Beyond the paradigm of active Brownian particles with rotational
diffusion in two dimensions, one could also explore three-dimensional
particles, stochastic variations in the propulsion speed as in active
Ornstein-Uhlenbeck models~\cite{magg15,fara15,fodo16}, or sudden
reorientations of the propulsion direction as in run-and-tumble
models~\cite{cate13,solo15}. Our definitions of quantities
characterising the energetic performance of engines apply already to
these cases, thus supporting the generality of our approach. Yet, new
design principles for the optimisation of the performance may emerge
in such more complex settings. It will also be interesting to
investigate the universality of the cooperative enhancement of the
performance beyond the mean-field approach used here. Thanks to modern
techniques for the microfabrication of particles~\cite{dile10,kuem13},
the exertion of forces using optical tweezers~\cite{blic12,kris16},
and the realisation of artificial self-propelled
particles~\cite{Golestanian2007,volp11}, it should be possible to
address these questions experimentally.

\begin{acknowledgments}
 Work funded in part by the European Research Council under the EU's
 Horizon 2020 Programme, Grant No.~740269. \'E.F. benefits from
an Oppenheimer Research Fellowship from the University of Cambridge, and a Junior
Research Fellowship from St Catharine’s College. M.E.C. is funded by the
 Royal Society. 
\end{acknowledgments}

\appendix

  \section{Limiting cases for the lattice model}
\label{sec:app_twostate}

In the limiting case of highly persistent active particles,
$\gamma\ll w_0,k_0$, there is a timescale separation between the
reorientations and the lateral transitions. Hence, the distribution
$p(i,n)$ can be written in terms of two effective Boltzmann
distributions $p(i,n)\approx \exp[-V_\mathrm{eff}(i,n)]/Z_n$ that are
normalised such that $\sum_i p(i,n)=1/2$ for each $n$.  The effective
potential must then obey
\begin{equation}
  V_\mathrm{eff}(i,n)-V_\mathrm{eff}(i+1,n)=\ln\frac{w_{i}^-+k_{i,n}^+}{w_{i+1}^++k_{i+1,n}^-}
  \label{eq:Veff}
\end{equation}
to restore a detailed balance relation for the combined transition
rates between adjacent states $i$. 
Note that even in this case, where the relative coordinate
equilibrates locally, the total system is nonetheless in a genuine
nonequilibrium state with nonvanishing currents $J$ and
$\Pa$. Provided that the lattice size $L$ is sufficiently large to separate the
attractive site from the repulsive one, lattice sites away from the
passive particle become depleted, such that the resulting currents no
longer depend on $L$.

In contrast, for $\gamma\gg w_0,k_0$, 
the orientation of $n$ equilibrates locally
for every $i$, leading to $n$-independent, effective transition rates
of the active particle of the form
\begin{equation}
k_{i,\mathrm{eff}}^\pm\equiv\frac{1}{2}(k_{i,+1}^\pm+k_{i,-1}^\pm)=k_0\cosh(\fa)\exp[(V_i-V_{i\pm 1})/2].
\end{equation}
Thus, the dynamics of $i_a$ and $i_p$ becomes equivalent to a passive
system that is driven only by the external force. Without an external
force, the system then reaches an effective equilibrium state
$p(i)\propto\exp(-V_i)$ with the actual interaction potential and
vanishing current $J$, which is analogous to the small persistence
regime in continuous models, where effective Boltzmann approaches are
legitimate~\cite{solo15,fodo18}. As leading-order corrections for fixed $k_0$ and $w_0$,
the deviations from the Boltzmann distribution, the current $J$ in Eq.~\eqref{eq:disc_current}, and
the optimal external force all scale like $1/\gamma$ (similar to the
case of off-lattice models~\cite{doer94}), leading to the
maximum extracted power scaling like $1/\gamma^2$.  Since, nonetheless, the
active power~\eqref{eq:pac_def} remains finite, the active efficiency
vanishes like $1/\gamma^2$ as well.

\begin{figure}
  \centering
  \includegraphics[width=\textwidth]{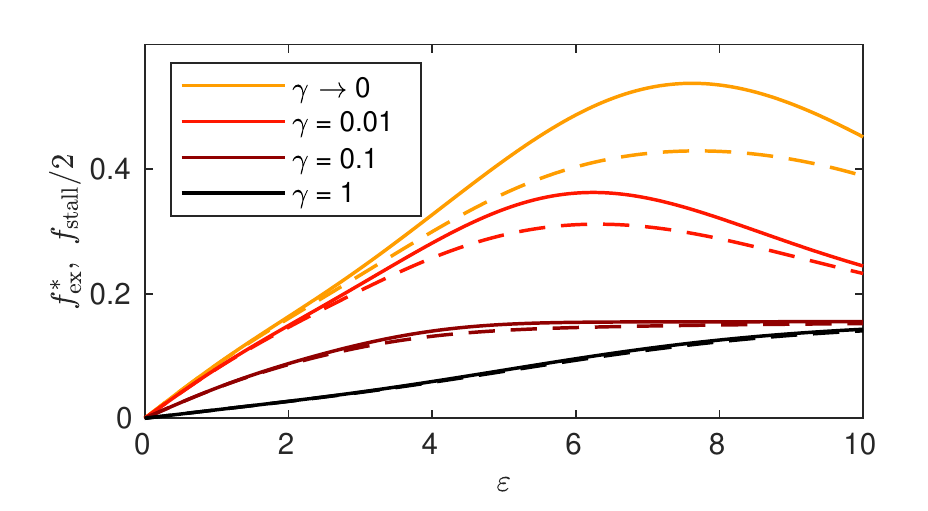}
  \caption{Optimal external force $\fe^*$ (solid lines) leading to the
    maximal extracted power (shown in the bottom panel in
    Fig.~\ref{fig:maxpow}. For small external forces within linear
    response, this force is half the stall force $f_\mathrm{stall}$ at
    the respective parameters (dashed lines); deviations occur for
    stronger forces.}
  \label{fig:maxpow_b}
\end{figure}

For small external forces, the response in the change of the
current must be linear. For a small asymmetry~$\varepsilon$ in the
interaction potential or large $\gamma$, the stall force $\fs$ is
small as well, such that the linear regime covers $\fs$. In this case, the
output current is given by $J=J_0(1-\fe/\fs)+\mathcal{O}(\fe^2)$ with
the current $J_0$ at zero force, such that the maximal extracted power
$P_\mathrm{max}=J_0\fs/4$ is attained at the force
$\fe^*=\fs/2$. Figure~\ref{fig:maxpow_b} compares these two
characteristic forces.

Another limiting case that can be understood analytically is the one
for which the interaction potential is strong,
i.e. $\varepsilon\to\infty$. In this limit, the stationary probability
is concentrated in the state $i=1$, which is almost impossible to
leave. Thus, there are almost no transitions and both the extracted
power and the active power vanish. Nonetheless, the active
efficiency~\eqref{eq:eta_ac} is well defined in this limit. It can be
calculated from the dominant contributions to the currents stemming
from rare and short-lived excursions to the state $i=2$. Jumps out of
this state are highly biased towards $i=1$, such that all other states
$i>2$ can be neglected for the calculation of the current. We assume a timescale separation between
the sojourn time in the state $i=2$ and the much larger timescale
$1/\gamma$ for reorientations. The steady-state distribution then
reads
\begin{align}
  p(1,\pm)&=\frac{1}{2}\frac{k_{2,\pm}^-+w_2^+}{k_{1,\pm}^++w_1^-+k_{2,\pm}^-+w_1^-},\nonumber\\ p(2,\pm)&=\frac{1}{2}\frac{k_{1,\pm}^++w_1^-}{k_{1,\pm}^++w_1^-+k_{2,\pm}^-+w_1^-}.
\end{align}
The stationary currents due to jumps of the passive particle follow
as 
\begin{equation}
  J_\pm= p(2,\pm)w_2^+- p(1,\pm)w_1^-=\frac{1}{2}\frac{k_{1,\pm}^+w_2^+-k_{2,\pm}^-w_1^-}{k_{1,\pm}^++w_1^-+k_{2,\pm}^-+w_1^-},
\end{equation}
which leads to the output current $J=J_++J_-$ and the active
efficiency
\begin{equation}
  \eta=\frac{\fe}{\fa}\frac{J_++J_-}{J_+-J_-}.
\end{equation}
Using the explicit forms for the rates, 
\begin{align}
w_1^-&=w_0\mre^{(\fe-\varepsilon)/2},\quad
w_2^+=w_0\mre^{-(\fe-\varepsilon)/2},\quad\nonumber\\ 
k_{2,\pm}^-&=k_0\mre^{-(\pm\fa-\varepsilon)/2},\quad
k_{1,\pm}^-=k_0\mre^{(\pm\fa-\varepsilon)/2},
\end{align}
we obtain
\begin{equation}
  J_\pm=\frac{1}{2}\frac{w_0 k_0\sinh\left(\frac{\pm\fa-\fe}{2}\right)}{k_0\cosh\frac{\pm\fa-\varepsilon}{2}+w_0\cosh\frac{\fe-\varepsilon}{2}}.
\end{equation}
Thus, as expected, both currents
vanish in the limit $\varepsilon\to\infty$. Nonetheless, $\eta$ remains finite in this limit. Additionally taking the limit $w_0\to 0$
yields 
\begin{equation}
  \lim_{w_0\to
    0}\lim_{\varepsilon\to\infty}\eta=\frac{\fe}{\fa}\frac{\cosh{\fa}-\exp\fe}{\sinh\fa}.
\label{eq:etalimit}
\end{equation}
independently of the order in which the limits are taken.
For large $\fa$ and $\fe\sim \fa-\sqrt{\fa}$, this efficiency gets
arbitrarily close to one, showing that there is no universal upper
bound on the efficiency smaller than $\eta\leq 1$.

\section{Implementation of the hard-core interaction}
\label{sec:hardcore}
We calculate numerical solutions of the Langevin
equation~\eqref{eq:langevin_rel} for the relative coordinate $\vec{r}$
in the limit of vanishing noise terms $\vec{\zeta}_a$ and $\zeta_p$
and the interaction potential $V(\vec{r})$ being hard core.
Away from the obstacle, the equations is integrated exactly using a
simple Euler scheme with time step $\delta t=0.01$ and zero potential
force. If it is detected that after the next such time step the active
particle would penetrate the obstacle, the equation is modified to 
\begin{equation}
  \dot{\vec{r}}=\vec{v}^0(\theta)+(\mu_a\vec{1}+\mu_p\vec{e}_x\otimes\vec{e}_x)\vec{f}_c.
\end{equation}
Therein, the effect of the potential force is modelled by the
constraint force $\vec{f}_c=f_c\vec{m}$, which has to be parallel to
the local normal vector to the surface of the obstacle $\vec{m}$. The
absolute value of the constraint force is obtained by requiring that
the resulting velocity is parallel to the surface
($\dot{\vec{r}}\cdot\vec{m}=0$), leading to
\begin{equation}
  f_c=-\frac{\vec{m}\cdot\vec{v}^0}{\mu_a+\mu_p(\vec{m}\cdot\vec{e}_x)^2}.
\end{equation}
If it is detected that the active particle has reached a cusp node of
the obstacle where it gets trapped, the calculation terminates and the
averaged relative velocity $\vec{v}(\theta)$ is assigned zero. 

We stress that the explicit evaluation of the average
$\mean{\vec\nabla V(\vec{r})}_\theta$ of the hard core potential
can be avoided by using Eq.~\eqref{eq:Vcomponents}, which uses only
average relative velocities.

\bibliography{../activework}

\end{document}